\documentclass[11pt,letterpaper]{article}
\usepackage{jheppub}
\usepackage{latexsym,amsmath,amsfonts,amssymb}
\usepackage[latin1]{inputenc}
\usepackage{graphicx}
\usepackage{times}

\usepackage{euscript,mathrsfs}
\usepackage{bbm}
\usepackage[american]{babel}
\usepackage{verbatim}

\newcommand{\be}{\begin{equation}}
\newcommand{\ee}{\end{equation}}
\newcommand{\bea}{\begin{eqnarray}} 
\newcommand{\eea}{\end{eqnarray}}

\newcommand{\bra}[1]{\langle#1|} \newcommand{\ket}[1]{|#1\rangle}
\newcommand{\ketbra}[2]{|#1\rangle\!\langle#2|}

\newcommand{\smfrac}[2]{\mbox{$\frac{#1}{#2}$}} \newcommand{\tr}{\operatorname{tr}}
\newcommand{\ox}{\otimes} \newcommand{\dg}{\dagger} 
\newcommand{\cA}{{\cal A}}  
  
 \newcommand{\cH}{{\cal H}}

\newcommand{\rar}{\rightarrow}

\newcommand{\pair}[2]{{\langle #1, #2 \rangle}}
\def\>{\rangle}
\def\<{\langle}

\def\a{\alpha}

\def\g{\gamma}
\def\d{\delta}
\def\e{\epsilon}

\def\s{\sigma}

\def\ps{\psi}
\def\o{\omega}

\def\X{\Xi}

\newcommand{\ZZ}{{{\mathbb Z}}}

\newcommand{\nn}{\nonumber}

\renewcommand{\a}{\alpha}

\renewcommand{\o}{\omega}

\newcommand{\Z}{\mathbb{Z}}

\title{Towards the fast scrambling conjecture}

\author[a]{Nima Lashkari}
\author[b]{Douglas Stanford}
\author[c,d]{Matthew Hastings}
\author[e]{Tobias Osborne}
\author[a,f,g]{Patrick~Hayden}

\emailAdd{lashkari@physics.mcgill.ca}
\emailAdd{salguod@stanford.edu}
\emailAdd{xhastings@gmail.com}
\emailAdd{tobias.j.osborne@gmail.com}
\emailAdd{patrick@cs.mcgill.ca}

\affiliation[a]{Department of Physics, McGill University, Montreal, QC, Canada}
\affiliation[b]{Stanford Institute for Theoretical Physics, Department of Physics, Stanford University, Stanford, CA, USA}
\affiliation[c]{Department of Physics, Duke University, Durham, NC, USA}
\affiliation[d]{Microsoft Station Q, Santa Barbara, CA, USA}
\affiliation[e]{Institut f\"ur Theoretische Physik, Hannover, Germany}
\affiliation[f]{School of Computer Science, McGill University, QC, Canada}
\affiliation[g]{Perimeter Institute for Theoretical Physics, Waterloo, ON, Canada}

\abstract{Many proposed quantum mechanical models of black holes include highly nonlocal interactions. The time required for 
thermalization to occur in such models should reflect the relaxation times associated with classical black holes in general 
relativity. Moreover, the time required for a particularly strong form of thermalization to occur, sometimes known as scrambling,
 determines the time scale on which black holes should start to release information. It has been conjectured that black holes 
scramble in a time logarithmic in their entropy, and that no system in nature can scramble faster. In this article, we address 
the conjecture from two directions. First, we exhibit two examples of systems that do indeed scramble in logarithmic time: 
Brownian quantum circuits and the antiferromagnetic Ising model on a sparse random graph. Unfortunately, both fail to be truly ideal fast scramblers for
 reasons we discuss. Second, we use Lieb-Robinson techniques to prove a logarithmic lower bound on the scrambling time of systems
 with finite norm terms in their Hamiltonian. The bound holds in spite of any nonlocal structure in the Hamiltonian, which might
 permit every degree of freedom to interact directly with every other one.}


\begin{document}
\maketitle

\section{Introduction}\label{intro}
There is a growing consensus based on evidence from string theory and gauge-gravity
correspondences that black holes do not destroy information when
they evaporate. Roughly, the argument is that black holes can be realized in string theory in a manner that accounts for their entropy~\cite{Susskind:1993ws,Sen:1995in,strominger:1996a,Callan:1996dv,Das:1996ug,Maldacena:1996ds,Horowitz:1996nw}, and that certain string theories are equivalent to manifestly unitary systems \cite{Banks:1996vh,maldacena:1997a,maldacena:2003a}. For a recent review, see \cite{balasubramanian:2011a}. 

Instead of being lost, information about the microscopic state of
the black hole leaks out with the hole's Hawking radiation, much as
it would for any other radiating object. Early estimates for the
amount of time it would take to recover a bit from a black hole,
however, suggested that no information would leak out for an amount
of time proportional to the black hole lifetime~\cite{page:1993a,page:1993b,susskind:1993a}. Since astrophysical
black holes have lifetimes many orders of magnitude longer than the
age of the universe, that is tantamount to the information being
lost forever. More specifically, such a long delay before the escape
of information provided a plausible resolution to some of the
conceptual conundrums of quantum gravity, most notably the apparent
inconsistency of information release with the quantum no-cloning
principle~\cite{susskind:1993a}.

More recent estimates using techniques from quantum information
theory, on the other hand, suggest that information could be
released from black holes much more quickly~\cite{hayden:2007a}. Those calculations
indicate that the relevant time scale is not the amount of time it
takes for the black hole to evaporate but, instead, the amount of
time the dynamics takes to ``scramble'' the black hole's microscopic degrees of freedom in
such a way that initially localized perturbations become
undetectable by observables that fail to probe a significant fraction of \emph{all} the degrees of freedom.
While a direct calculation of
this scrambling time remains out of reach, the relaxation timescales
associated with classical black holes are incredibly fast. So fast,
in fact, that if they also govern the scrambling time, then the
black hole complementarity principle, one of the guiding principles for
many researchers in quantum gravity~\cite{susskind:1993a,kiem:1995a,lowe:1995a} is only just saved from
inconsistency -- faster scrambling would lead to a paradox.

Motivated by these considerations, as well as the implications of
the existence of fast scramblers for the underlying structure of the
degrees of freedom of quantum gravity, Sekino and Susskind
elaborated on the speculations of~\cite{hayden:2007a} to formulate the 
following three-part fast scrambling conjecture~\cite{sekino:2008a,susskind:2011a}:
\begin{enumerate}
\item The most rapid scramblers take a time logarithmic in the
number of degrees of freedom.
\item Matrix quantum mechanics (systems whose degrees of freedom are
$n$ by $n$ matrices) saturate the bound.
\item Black holes are the fastest scramblers in nature.
\end{enumerate}
The purpose of this article is to explore the validity of the
conjecture, focusing primarily on the first part. While the
conjecture implicitly refers to the most rapid scramblers \emph{in
nature}, we allow ourselves the freedom to investigate the most
rapid scramblers \emph{in quantum mechanics} (and even slightly
beyond) without worrying if our models are physically realizable.

Thanks to earlier research in quantum computation by Dankert
\emph{et al.}, it is already known how to define a time-dependent
Hamiltonian which will scramble in logarithmic time with high
probability~\cite{dankert:2009a}. The scrambler, however, is a very carefully
engineered quantum circuit, so that it is difficult to ascribe the
fast scrambling specifically to interactions between the
constituents as opposed to clever tuning of their external knobs.
Ideally, therefore, we would like to exhibit a fast scrambler
described by a simple time-independent Hamiltonian. To that end, we
present two examples:
\begin{itemize}
\item {\bf Brownian quantum circuits.} The scrambler of \cite{dankert:2009a} was a
highly structured quantum circuit. Other work has studied circuits
composed of random gates~\cite{emerson:2005a,harrow:2009a,arnaud:2008a,brown:2010a,diniz:2011a} but a rigorous proof that they scramble in
logarithmic time remains to be found. Instead, we present a
continuous-time analog of a quantum circuit in which the Hamiltonian
is a stochastically varying two-body interaction, and prove that it
scrambles in logarithmic time.
\item {\bf Ising model.} We consider scrambling by the antiferromagnetic Ising
interaction on a general graph with an external field parallel to the spin quantization axis. 
Despite its triviality, this model nonetheless exhibits a form of weak
scrambling in logarithmic time on some
graphs.
\end{itemize}
The careful reader will have observed that neither of these
examples meets all of our criteria for a convincing scrambler: the
Brownian quantum circuits are time-dependent, if not structured, and the Ising model fails
to scramble fully. Nonetheless,
we feel that, taken together, the examples provide substantial
evidence that quantum systems with simple time-independent
Hamiltonians \emph{can} scramble in logarithmic time.

The fast scrambling conjecture not only states that logarithmic-time
scramblers exist, but also asserts that it is
impossible to scramble faster. It might seem
hopeless to address this question without invoking additional
physical assumptions beyond just the validity of quantum mechanics.
After all, scrambling is a form of information propagation, and
limits on information propagation normally depend on
locality. A Hamiltonian allowing all degrees of freedom to interact directly has no
locality to speak of. Nonetheless, using bounds of
Lieb-Robinson-type~\cite{lieb:1972a,nachtergaele:2006a,hastings:2006a} to rigorously control a mean-field approximation, we are able to show the following:
\begin{itemize}
\item Subject to some nontrivial norm assumptions on the terms in
the Hamiltonian, no physical system described by a Hamiltonian with dense
two-body interactions can scramble in time faster than
$O({\log n})$, where $n$ is the number of degrees of freedom.\footnote{Throughout the article, $O(f(n))$ is used in the physicist's sense of ``leading order''. Readers familiar with asymptotic notation should for the most part reinterpret these expressions as $\Theta(f(n))$.}
``Dense'' here means that the number of interacting pairs of degrees of freedom scales like $O(n^2)$.
\item The bound extends to certain four-body
Hamiltonians similar to the BFSS matrix model~\cite{Banks:1996vh}.
\item With more sparsely interacting systems, there is a lower bound of $O(\sqrt{\log n})$ on the scrambling time.
\end{itemize}
While the norm assumptions are unfortunately too stringent to allow
us to apply the results rigorously to the matrix model Hamiltonian and,
thereby, to black hole physics, these results are strong evidence
that scrambling in less than logarithmic time is impossible. (A related obstacle is
our focus on distinguishable degrees of freedom; bosonic degrees of freedom naturally lead to unbounded operators.)

\subsection{Related work}

Asplund, Berenstein and Trancanelli \cite{Asplund:2011qj} have numerically investigated relaxation in matrix models. Their approach is to look at the classical dynamics of the system, with initial states selected stochastically in such a way as to enforce the uncertainty principle. They do indeed find what appears to be very rapid relaxation of the system to an attractor state, but their article only considers a fixed-sized and relatively small system, so it cannot directly address the scaling of relaxation time with system size. The relationship between this classical relaxation time and quantum mechanical scrambling is also an interesting and currently unexplored question.

Barbon and Magan \cite{Barbon:2011pn} have approached the conjecture from a different direction. They suggest that the logarithmic factor in the black hole scrambling time arises from the hyperbolic geometry of the so-called ``optical metric'' $ds^2/g_{00}$ associated to a simple coordinatization of Rindler space. Specifically, they argue that the Lyapunov time for a classical billiards game on such a geometry agrees with the scrambling time.

More indirectly, while most work prior to \cite{hayden:2007a} argued that black holes held information for an amount of time comparable to the black hole lifetime, if not forever, occasional hints were found that information might leak out faster~\cite{schoutens:1993a}. Reversing the reasoning, one could interpret such arguments as evidence in favour of the fast scrambling conjecture.

The seemingly paradoxical idea that a closed quantum system undergoing unitary dynamics can exhibit {equilibration} or {thermalization} is an old one dating back, at least, to von Neumann~\cite{vonneumann:1929a}; the apparent contradiction with the fact that the global state is pure and never equilibrates is resolved by noticing that any small subregion in an interacting closed quantum system generically becomes {entangled} with the rest and may appear, at least {locally}, thermal. For large systems the recurrence time is extremely long so, for all intents and purposes, it is meaningful to say that the system has become (locally) thermalized.  There is now an enormous literature on this topic (see e.g., \cite{mahler:2009a} for a textbook treatment). Recently these old questions have received new impetus from quantum chaos, quantum information theory, and many-body physics, all of which have brought new tools to bear \cite{linden:2009a, goldstein:2006a, popescu:2006a, reimann:2008a, bocchieri:1959a, lloyd:1988a, tasaki:1998a, calabrese:2006a, deutsch:1991a, srednicki:1994a, riera:2011a, rigol:2011a} leading to an emerging understanding of the general conditions under which a closed quantum system will exhibit (local) thermalization.

\section{Scrambling: definition and properties} \label{sec:defn}

Scrambling is nothing other than a strong form of thermalization applicable to closed system evolution. A closed system never forgets its initial state, but over time it might become impossible to distinguish different initial states without measuring a large fraction of all the system degrees of freedom. The minimum time required for the information about the initial state to be lost is called the {\it scrambling time}.

In general, the scrambling time depends on the nature of the set of initial states. For example, small perturbations of an equilibrium configuration will generally get scrambled more rapidly than will a pair of metastable configurations. Likewise, it could be easier to scramble a discrete set of states than all possible superpositions of those states.  In this article, we will focus on product initial states, but a slightly different formulation will likely be necessary in order to study black hole physics. In particular, energy conservation will usually prohibit the strong form of scrambling we demand here.\footnote{A general definition of scrambling appropriate to finite temperature will be included in an upcoming revision of this article.}

Suppose that we have a system with $n$ distinguishable degrees of freedom  
and a Hamiltonian
$
H = \sum_{\langle x, y \rangle} H_{\langle x, y \rangle}
$
acting on a Hilbert space $\cH = \cH_1 \ox \cH_2 \ox \cdots \ox \cH_n$,
where the sum ranges over pairs $\langle x,y \rangle$ of degrees of freedom. 
%
%
An initial state $\ket{\Psi(0)}$ evolves to a state $\ket{\Psi(t)} = \exp(-iHt) \ket{\Psi(0)}$. For $S \subseteq \{1,2,\ldots,n\}$ a subset of the degrees of freedom and $S^c$ the complement,  let $\Psi_S(t) = \tr_{S^c} \ketbra{\Psi(t)}{\Psi(t)}$. 

Ideally, a scrambler will delocalize any  information initially localized with respect to the factorization of $\cH$ into subsystems. We therefore define the scrambling time $t_*$ to be smallest time $t$ such that $\Psi_S(t)  \simeq \Phi_S(t)$  for all $S$ such that $|S| < \kappa n$ for some $0 < \kappa < 1/2$, and for all initial states $\ket{\Psi(0)}$ and $\ket{\Phi(0)}$ that factorize into the form $\ket{\o_1} \otimes \ket{\o_2} \otimes \cdots \otimes \ket{\o_n}$. For concreteness, we will fix $\kappa = 1/3$, but its specific value will not affect our conclusions.

The scrambling time obviously depends on the normalization of the Hamiltonian. In Sekino and Susskind's original formulation, the fast scrambling conjecture was that $t_*/\beta \geq C(\beta) \log n$, where $\beta$ is the inverse temperature and $C$ is an unspecified function. In much of what follows, we will work either far from equilibrium, where $\beta$ is not be well-defined, or near infinite temperature, where it doesn't accurately reflect the energy per degree of freedom (which stays finite as $\beta \rightarrow 0$ in the spin models we consider). This leaves a couple of alternatives for a dimensionless measure of scrambling time:
\begin{itemize}
\item One can consider the {ratio} of the amount of time it takes to scramble systems of different sizes, hopefully cancelling the temperature dependence.
Let $t_*^{(k)}$ be the scrambling time for subsystems of size $|S| \leq k$ and set $\tau_* = t_*^{(\kappa n )} / t_*^{(1)}$. The revised conjecture is then that $\tau_* \geq O( \log n )$. 
\item 
The Hamiltonians we consider do not have their interactions arranged in a lattice structure. Instead, each subsystem $S$ generally participates in a number of interactions growing with $n$. As a second option, one can require that the energy scales extensively with the system size $n$, thereby selecting a normalization for the Hamiltonian which, while coarse, is sufficient to determine the scaling of $t_*$ with $n$.
\end{itemize}
The final step in formalizing the notion of scrambling time is to clarify the meaning of $\Psi_S(t) \simeq \Phi_S(t)$. The \emph{trace distance} provides a notion of statistical distinguishability that meshes well with the quantum information theoretic applications of scrambling. Specifically, one should demand that $\| \Psi_S(t) - \Phi_S(t) \|_* < \epsilon$ where $\| X \|_* = \tr \sqrt{X^\dg X}$. (See, \emph{e.g.}, \cite{mikeandike} for a discussion of the statistical interpretation of the norm.)

\subsection{Scrambling as entanglement generation} \label{subsec:scrambling.entanglement}

Scrambling information is by definition just storing that information in complicated correlations between many subsystems, which means that scrambling is intimately related to the production of entanglement. In fact, the concepts are essentially one and the same.
Intuitively, the reason is that if the restriction $\Psi_S(t_*)$ of a scrambled state is not highly mixed, then there won't be enough room in the Hilbert space $\cH$ at time $t_*$ to accommodate all the scrambled states, which contain a basis for $\cH$. (The relationship is simplest when $\cH$ is finite dimensional, which we will assume here but not elsewhere in the article.)

Formalizing that intuition is a simple exercise in quantum information theory.
Recall that  the von Neumann entropy of a density operator $\rho$ restricted to subsystem $A$ is $H(A)_\rho = H(\rho_A) = - \tr \rho_A \log \rho_A$, and that the mutual information between subsystems $A$ and $B$ for $\rho$ is defined as $I(A:B)_\rho = H(A)_\rho + H(B)_\rho - H(AB)_\rho$.

Fix an orthonormal product basis $\{ \ket{\psi_{x_1}} \ket{\psi_{x_2}} \cdots \ket{\psi_{x_n}} \}$ for $\cH = \cH_1 \ox \cH_2 \ox \cdots \ox \cH_n$. After time $t_*$, all of these product states will be scrambled, so consider $\ket{\Psi^{(x_1,\ldots,x_n)}} = \exp(-iHt_*) \ket{\psi_{x_1}} \otimes \cdots \otimes \ket{\psi_{x_n}}$. It it convenient to introduce an auxiliary Hilbert space $X$ and consider the following density operator on the combined $X\cH$ system:
\begin{equation}
\rho_{X \cH} = \frac{1}{\dim \cH} \sum_{x_1,\ldots,x_n} \ketbra{x_1,\ldots,x_n}{x_1,\ldots,x_n}_X \ox \Psi^{(x_1,\ldots,x_n)}.
\end{equation}
The system $X$ records in an orthonormal basis which state describes $\cH$, and the overall state is an equal mixture over choice of $x_1,\ldots,x_n$.

Because subsystem $S$ is scrambled, all of the states $\Psi^{(x_1,\ldots,x_n)}_S = \tr_{S^c} \Psi^{(x_1,\ldots,x_n)}$ will be essentially indistinguishable, so there can't be any significant correlations between $X$ and $S$. A quantitative way of expressing that fact is that the mutual information $I(X:S)_\rho$ will be small, say less than $\delta$. (A standard continuity result implies that $\delta$ can be chosen to be $3 \e \log \dim \cH + f(\e)$, where $f(\e)$ goes to zero with $\e$ and is independent of $n$~\cite{fannes:73a}.)

On the other hand, the states $\ket{\Psi^{(x_1,\ldots,x_n)}}$ form an orthonormal basis for $\cH$, so their equal mixture is just the maximally mixed state on $\cH$. The state $\rho_\cH$ is by construction precisely that equal mixture. It follows that $\rho_S$ is also maximally mixed and, therefore, that $H(S)_\rho = \log \dim \cH_S$.

Substituting into the inequality $I(X:S) < \delta$ then gives
\begin{equation} \label{eqn:big.conditional.h}
\log \dim \cH_S - \d < H(XS)_\rho - H(X)_\rho.
\end{equation}
The quantity on the righthand side, $H(XS)_\rho - H(X)_\rho$ is known as the conditional entropy $H(S|X)_\rho$ of $S$ given $X$. It can be interpreted as the uncertainty remaining in $S$ once $X$ is known and evaluates in this case to 
\begin{equation}
\frac{1}{\dim \cH} \sum_{x_1,\ldots,x_n} 
	H\left( \Psi_S^{(x_1,\ldots,x_n)} \right),
\end{equation}
the average entropy of the states $\Psi_S^{(x_1,\ldots,x_n)}$. Inequality~(\ref{eqn:big.conditional.h}) thus ensures that the states $\Psi_S^{(x_1,\ldots,x_n)}$ have high entropy, very close, in fact, to the maximum possible value of $\log \dim \cH_S$. (In the finite temperature setting, $\log \dim \cH_S$ would be replaced by the entropy of the appropriate thermal state on $S$.)

The entropy of a mixed state on $S$ measures how much entanglement there is between $S$ and $S^c$ in the corresponding pure state. Good scrambling can therefore only be achieved by a time evolution that produces nearly maximal entanglement, and vice versa. 

\section{Brownian quantum circuits} \label{sec:Brownian}

A quantum circuit is an idealized model of the time evolution of a quantum computer, which is generally assumed to consist of a number of qubits. At a given discrete time step, a collection of ``gates'' is applied to the state, where a gate is a unitary transformation involving one or two qubits. Each qubit participates in at most one gate per time step. 

As mentioned earlier, Dankert \emph{et al.} found a quantum circuit that scrambles $n$ qubits after $O(\log n)$ time steps~\cite{dankert:2009a}. Their circuit, however, is quite an intricate construction that doesn't plausibly model any naturally occurring interactions. Other researchers have studied random quantum circuits, establishing that they are scramblers, but the question of whether they scramble in time $O(\log n)$ remains open~\cite{emerson:2005a,harrow:2009a,arnaud:2008a,brown:2010a,diniz:2011a}.

In this section, we study a continuous-time analog of a random quantum circuit, which provably does scramble in time $O(\log n)$. Consider $n$ qubits interacting according to a stochastically varying Hamiltonian. Time is subdivided into steps of length $\epsilon = \Delta t$ and during a given time step, the interaction between each pair of qubits is given by a random Wigner matrix. More formally, 
the Hamiltonian from time $t_r = r \Delta t$ to $t_{r+1} = (r+1)\Delta t$ is given by
\begin{equation}
H_r = \sum_{j < k} \sum_{\a_j,\a_k = 0}^3 \sigma_j^{\a_j} \otimes \sigma_k^{\a_k}  \; \Delta B_{r,j,k,\alpha_j,\alpha_k} ,
\end{equation}
where the $\Delta B_{r,j,k,\alpha_j,\alpha_k}$ are independent and identically chosen real Gaussians $N(0,\epsilon^2)$ with zero mean and variance $\epsilon^2$ . The operator $\sigma_j^{\a_j}$ represents the Pauli operator $\sigma^{\alpha_j}$ acting on qubit $j$, with $\sigma^0$ the identity matrix. 

The time evolution from $t_0$ to $t_r$ is given by
\begin{equation}
\exp( -i H_{r-1} \Delta t ) \exp( -i H_{r-2} \Delta t ) \cdots \exp( -i H_0 \Delta t).
\end{equation}
For this process to have a well-defined and nontrivial limit as $\Delta t \rar 0$, one must choose $\epsilon^2 \propto (\Delta t)^{-1}$~\cite{karatzas:1991a}. That is, the strength of the interactions must increase as the size of the time steps decreases. This requirement makes it problematic to interpret $t_*$ in units of energy. Instead, we show that the ratio $\tau_* = t_*^{(\kappa n)} / t_*^{(1)} = O( \log n )$ for constant $0 < \kappa < 1/2$.
More generally, the ratio of the time required to scramble systems of size $k$ to the time required to scramble a single qubit scales like $O( \log k )$. 


%
The limiting dynamics of the random Hamiltonian evolution is given by $U(0) = I$ and $U( t + dt ) = \exp( i \, dG(t) ) \, U(t)$ for
\begin{equation}
dG(t) =
\frac{1}{\sqrt{8n(n-1)}}\sum_{j<k}\sum_{\alpha_j,\alpha_k=0}^3
\sigma_j^{\alpha_j}\otimes\sigma_k^{\alpha_k} \: dB_{j,k,\alpha_j,\alpha_k}(t),
\end{equation}
where the $dB_{j,k,\alpha_j,\alpha_k}(t)$ are independent Brownian motions with unit variance per unit time. Since we are only interested in $\tau_*$, the normalization factor is of no real consequence; it is chosen 
such that $\|dG(t)\|^2_2=dt$.

Calculating using the Ito calculus (see~\cite{arnold:1974a} for an accessible introduction) leads to the following stochastic differential equation for $U(t)$:
\begin{equation}
dU(t) =
\frac{i}{\sqrt{8n(n-1)}}\sum_{j<k}^n\sum_{\alpha_j,\alpha_k=0}^3
\sigma_j^{\alpha_j}\otimes\sigma_k^{\alpha_k}\otimes I_{\setminus\{j,k\}}\:U(t)\: dB_{\alpha_j,\alpha_k}(t)
- \frac12U(t)dt.
\end{equation}
(In a slight abuse of notation, we henceforth write $dB_{\a_j,\a_k}(t) := dB_{j,k,\alpha_j,\alpha_k}(t)$. $I_{\setminus\{j,k\}}$ denotes the identity on all sites except for $i$ and $j$.)
Suppose we have some initial state $|\Psi(0)\rangle$. Then the state
$\Psi(t) = U(t)|\Psi(0)\rangle\langle \Psi(0)|U^\dag(t)$ undergoes
the dynamics
\begin{multline}\label{dpsi}
d\Psi(t) = \frac{i}{\sqrt{8n(n-1)}}\sum_{j<k}\sum_{\alpha_j,\alpha_k=0}^3
[\sigma_j^{\alpha_j}\otimes\sigma_k^{\alpha_k}\otimes I_{\setminus\{j,k\}}, \Psi(t)]
dB_{\alpha_j,\alpha_k}(t) - \Psi(t) dt + \\ \frac{1}{8n(n-1)}\sum_{j<k}
\sum_{\alpha_j,\alpha_k=0}^3
\left(\sigma_j^{\alpha_j}\otimes\sigma_k^{\alpha_k}\otimes I_{\setminus\{j,k\}}\right)\Psi(t)\left(\sigma_j^{\alpha_j}\otimes\sigma_k^{\alpha_k}\otimes I_{\setminus\{j,k\}}\right)
dt.
\end{multline}

The time evolution will have scrambled subsystem $S$ once $\Psi_S(t)$ is independent of the initial state, as measured by the trace distance as discussed in Section~\ref{sec:defn}. Equivalently, $\Psi_S(t)$ should approach a fixed state independent of $\Psi(0)$. In the case of Brownian circuits, that fixed state is close to maximally mixed provided $S$ is not too large. Rather than calculating $\| \Psi_S(t) - I_S / \dim \cH_S \|_1$ directly, it is much easier to evaluate
\begin{equation}
\| \Psi_S(t) - I_S / \dim \cH_S \|_2^2 = \tr \Psi_S(t)^2 - \frac{1}{\dim \cH_S}.
\end{equation}
An application of Cauchy-Schwarz ensures that if  $\tr \Psi_S(t)^2 < (1+\e^2)/\dim \cH_S$, then $\| \Psi_S(t) - I_S / \dim \cH_S \|_1 < \e$, as required for scrambling.

We therefore introduce the {\it purity} of a subsystem $S$:
\begin{equation}
h_{S}(t) \equiv \tr(\Psi_S(t)^2).
\end{equation}
The equation of motion for the purity $h_S(t)$ is given by
\begin{equation}\label{dyn}
dh_{S}(t) = 2\tr(\rho_S(t)d\rho_S(t)) + \tr((d\rho_S(t))^2).
\end{equation}
After some algebra, it is shown in Appendix \ref{BrownApp} that (\ref{dyn}) gives
 the following dynamics for the purity  averaged over realizations of the Brownian motion, $\overline{h}_S = \mathbb{E}_{B}[h_S]$:
\begin{multline}
n(n-1)\frac{d\overline{h}_{S}(t)}{dt} = 2 |S^c|\sum_{j\in
S} \overline{h}_{S\setminus \{j\}}(t) +
2\left(|S^c|(|S^c|-1) + |S|(|S|-1) -n(n-1)\right)\overline{h}_S(t) +  \\
2|S|\sum_{k\in S^c}\overline{h}_{S\cup\{k\}} -
\sum_{\substack{j\in S\\ k\in
S^c}}\overline{h}_{S\setminus\{j\}\cup\{k\}}.
\end{multline}
\begin{figure}
\begin{center}
\includegraphics[height=5cm]{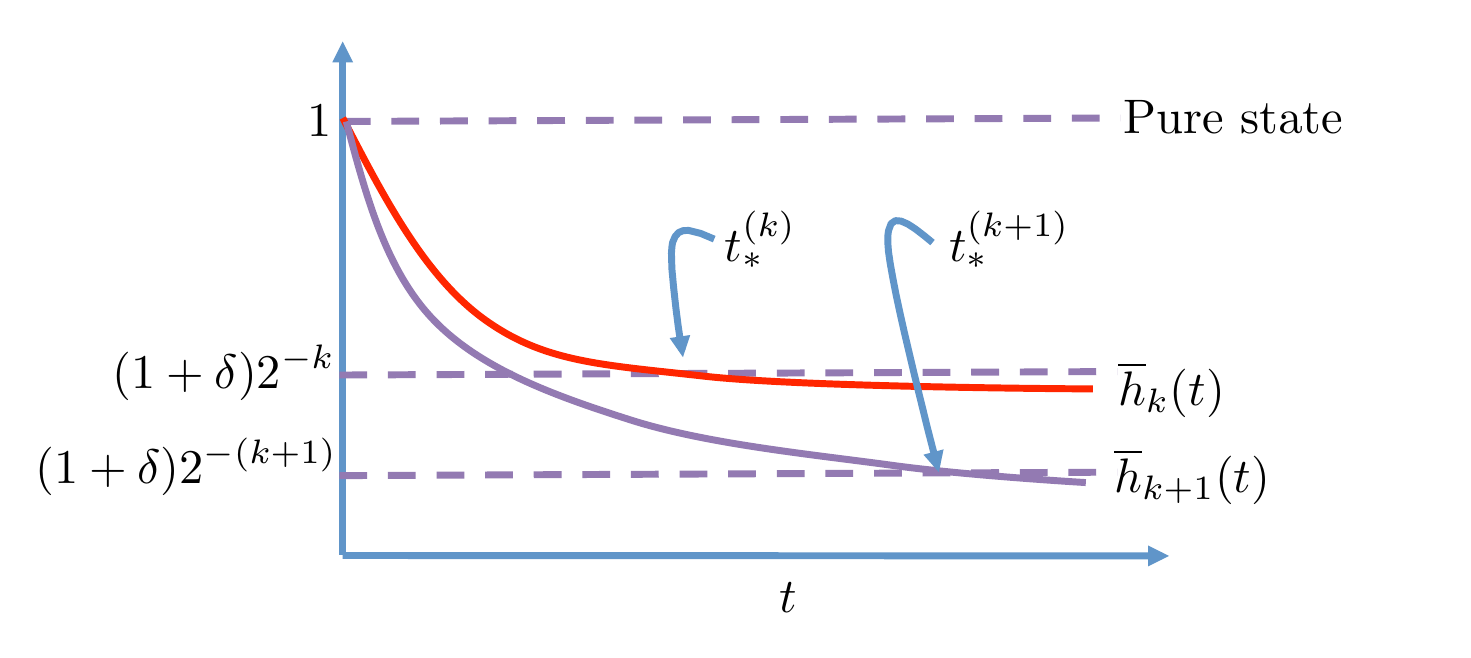}
\caption{Schematic plot of the decay of the average purity $\overline{h}_k(t)$ of a subsystem $S$ of size $k$. When the initial state is a pure product state all purities begin equal to one. The scrambling time for a system of size $k$ is defined as the amount of time required before purity of subsystems of size $k$ becomes less than $(1+\d)2^{-k}$; a purity of exactly $2^{-k}$ corresponds to the maximally mixed state. For subsystems of size smaller than $n/2$, the dynamics ensures that larger systems have smaller purities, a property not necessarily true of general entangled states.} \label{fig:purity.decay}
\end{center}
\end{figure}
Here $|A|$ means log dim $A$. If the initial configuration $\Psi(0)$ consists of a pure product state, then $\overline{h}_S$ depends only on $|S|=k$, so the system of ODE's collapses to a tridiagonal system and can be written in the form
\begin{equation}\label{ODE}
\frac{d\overline{h}_k(t)}{dt}=\frac{k(n-k)}{n(n-1)}\left(2\overline{h}_{k+1}+2\overline{h}_{k-1}-5 \overline{h}_k\right).
\end{equation}
The rough features of the system (\ref{ODE}) are sketched in Figure~\ref{fig:purity.decay} and the system's behavior is studied in Appendix \ref{ODEApp}, with the conclusion that the ratio scrambling time 
\begin{equation}
\tau_* = t_*^{\kappa n}/t_*^1 \sim \log n.
\end{equation}


\section{Ising interaction on random graphs} \label{sec:ising}

There is an inherent difficulty in searching for fast scramblers:
the intuition that a given system will rapidly scramble information
is usually based on a sense that the dynamics is complicated, which is
almost invariably an obstacle to studying the details of the system's time
evolution. Complexity is not an absolute requirement, however. In this section, we will see that
one of the simplest conceivable quantum mechanical systems has lessons to teach
us about scrambling time.

\begin{figure}
\begin{center}
\includegraphics[width=6cm]{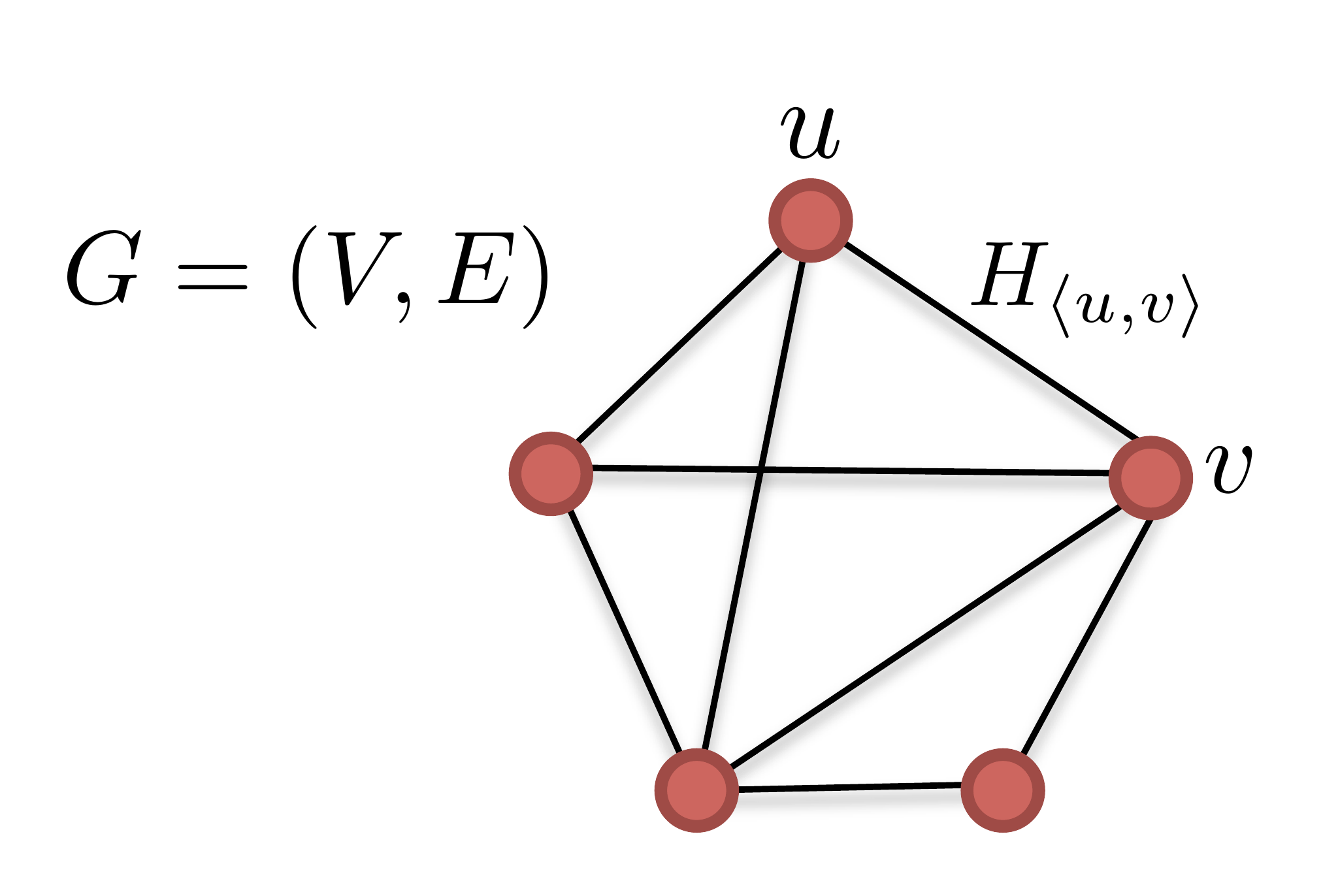}
\caption{Antiferromagnetic Ising interaction on an undirected graph $G=(V,E)$. There is term $H_{\langle u,v \rangle}$ in the Hamiltonian for each edge $\langle u,v \rangle \in E$ of the graph. Generic sparse graphs with average vertex degree roughly $\log |V|$ will quickly scramble information stored in the simultaneous $\{ \sigma^x_v : v \in V \}$ eigenbasis.} \label{fig:graph}
\end{center}
\end{figure}

Let $G=(V,E)$ be an undirected graph. Assign a spin-$\smfrac{1}{2}$
to each vertex $v \in V$ and allow spins adjacent with respect to
the edge set $E$ to interact via the antiferromagnetic Ising
Hamiltonian
\begin{equation}
H = \frac{|V|}{|E|} \sum_{\langle u,v \rangle \in E} \frac{1}{4} ( I - \s^z_u )
\ox (I - \s^z_v)
\end{equation}
as illustrated in Figure~\ref{fig:graph}.
The normalization factor $|V|/|E|$ is chosen to ensure that the
energy per spin scales extensively with the system size, $n = |V|$,
as discussed in Section \ref{sec:defn}. Choosing
$\ket{0_z}$ and $\ket{1_z}$ to be the $+1$ and $-1$ eigenstates of
$\sigma^z$, the Hamiltonian can be written more simply as
\begin{equation}
\label{isingh}
H = \frac{|V|}{|E|} \sum_{\langle u,v \rangle \in E} \ketbra{1^z}{1^z}_{u} \ox
\ketbra{1^z}{1^z}_v.
\end{equation}
The system obviously can't scramble because any product state of the
form $\ket{i_1^z}\ket{i_2^z}\cdots\ket{i_n^z}$ is an eigenstate of $H$.
Local information encoded in that basis remains locally accessible
for all times. On the other hand, information in the conjugate basis
of $\sigma^x$ eigenstates, $\ket{0^x}$ and $\ket{1^x}$, potentially
has more interesting behavior. Suppose then that the initial state
is $\ket{\Psi(0)} = \ket{i_1^x} \ket{i_2^x} \cdots \ket{i_n^x}$.

\newcommand{\Te}{t_{\operatorname{ent}}}

Up to a global phase, the system is periodic with period $\pi |E| / |V|$ and
the state $\ket{\Psi(t)}$ at time $t$ is most entangled at time $\Te
=\pi |E| / (2 |V|)$. The state $\ket{\Psi(\Te)}$ is known as a
\emph{graph state} in quantum computation, where it plays a central
role in the measurement-based quantum computing
architecture~\cite{raussendorf:2003a,nest:2006a}. For a subset $S \subseteq V$ of spins, the
entanglement entropy of the density operator $\Psi_S(\Te) = \tr_{S^c} \ketbra{\Psi(\Te)}{\Psi(\Te)}$ has a simple formula in
terms of the submatrix $\operatorname{Adj}_{S}$ of the adjacency
matrix of $G$ that selects the rows of $S$ and the columns of $S^c$~\cite{hein:2004a}:
\begin{equation} \label{eqn:Z2rank}
S( \Psi_S(\Te) ) = \operatorname{rank}_{\ZZ_2} \operatorname{Adj}_S,
\end{equation}
where the entropy is measured in bits. It follows that if
$\operatorname{Adj}_S$ has full rank as a matrix over $\ZZ_2$, then
the entanglement is $|S|$ bits. The only density operator with
$|S|$ bits of entropy on $|S|$ qubits, however, is the maximally
mixed density operator. Therefore, if $\operatorname{Adj}_S$ has
rank $|S|$, the final density operator on $S$ will be independent of
the choice of initial state $\ket{\Psi(0)} = \ket{i_1^x} \cdots
\ket{i_n^x}$. That is, the system will have scrambled the $\sigma^x$
eigenstates.

Each edge from $S$ to $S^c$ contributes a nonzero entry
to $\operatorname{Adj}_S$, but formula~(\ref{eqn:Z2rank}) implies that too many
connections can reduce entanglement. For example, for the fully
connected graph, every row of $\operatorname{Adj}_S$ is just a
sequence of ones, so there is never more than one bit of
entanglement entropy. To maximize the entanglement between $S$ and
$S^c$, one needs the matrix $\operatorname{Adj}_S$ to
have full rank for all $|S| \leq n/2$. This is generically the
case for appropriate random graphs in which edges are included randomly and
independently in $G$ according to the rule $\operatorname{Pr}[(u,v)
\in E] = p$. 

Since $\Te = \pi |E| / n$, minimizing $\Te$ requires minimizing the
expected number of edges in the graph, which is ${n \choose 2} p$,
subject to the constraint that the rank of $\operatorname{Adj}_S$ be 
maximal for all $|S| \leq n/2$. As $n$ goes to infinity and $\smfrac{|S|}{|S^c|}$ goes to any constant $\a$, the rank
defect of the matrix is Poisson distributed with parameter $\a e^{-\g}$ provided
$(\log n+\g)/n \leq p \leq 1 - (\log n+\g)/n$~\cite{kolchin:1999a}. Therefore,
$\operatorname{Adj}_S$ will be full rank with probability at least $1 -  e^{-\g}$.
Thus, the minimal value of $\Te$ is equal to $\pi(\log n+\g)$, where $\g$ can be regarded
as a constant.

Even though the system doesn't scramble fully in the sense of making
all local information locally inaccessible, it does scramble the
basis of $\sigma^x$ eigenstates and does so in time logarithmic in
$n$, as required of a fast scrambler.

For the sake of comparison with the Brownian circuit model, it is
also instructive to consider the analog of $\tau_*$, the ratio of
the amount of time to scramble systems of size $\kappa n$ to the time
required to scramble a single qubit. Since the system is exactly 
solvable, it is straightforward to establish by direct calculation that for $S = \{ j \}$ a
singleton, the Hamiltonian \eqref{isingh} and initial state $|\Psi(0)\rangle$ imply
\begin{equation}
\label{sspurity}
\text{tr}\rho_{\{j\}}^2(t) = \frac{1}{2} \left(1+\cos^{2d_j}\frac{n}{2|E|}t\right)
\end{equation}
where $d_j$ is the number of graph neighbors of site $j$. The expected number of neighbors per site is $p(n-1)$. Requiring that \eqref{sspurity} be close to minimal, \emph{i.e.} $\frac{1}{2}$, gives the 1-scrambling time as $O(\sqrt{pn}) = O(\sqrt{\log n})$. The ratio of the times required for scrambling
$\sigma^x$ eigenstates therefore scales like $O(\sqrt{\log n}) = O(\log n / \sqrt{\log
n} )$. This hints at the possibility that for systems that do scramble all product states, unlike this Ising model, 
$\tau_*$  might also fail obey an $\Omega(\log n)$ lower bound as required by
the fast scrambling conjecture. 

Regardless, the Ising model provides an example of a system capable of producing large scale multipartite entanglement sufficient to scramble all information stored locally in a fixed basis on a time scale no more than logarithmic with the number of degrees of freedom.

\section{Lower bounds on the scrambling time} \label{sec:lowerbounds}

One way to prove lower bounds on the scrambling time is to exploit the connection between scrambling and signalling. In particular, scrambling a subsystem $S$ implies the ability to signal to the complementary subsystem $S^c$. The main task of this section is therefore to prove signalling bounds, but we must do so without relying on relativity or, more generally, any underlying geometry in the organization of the degrees of freedom.
Our technique goes back to Lieb and Robinson~\cite{lieb:1972a}, who proved bounds on commutators $[ O_A(t) , O_B ]$ for observables $O_A$ and $O_B$ localized on subsystems $A$ and $B$ of lattice spin systems. To signal reliably from $B$ to $A$, there must be normalized observables for which the norm of the commutator is $O(1)$.
Hastings improved the original Lieb-Robinson technique so as to produce dimension-independent bounds~\cite{hastings:2004a} and Nachtergaele-Sims showed how to adapt it to general graphs~\cite{nachtergaele:2006a}. The version we start from combines both features and is due to Hastings and Koma~\cite{hastings:2006a}. 

As we will see, the Lieb-Robinson technique gives lower bounds on the time required to signal from $B$ to $A$ provided $A$ and $B$ are both constant-sized subsystems. The definition of scrambling used in this paper, however, only implies signalling from a constant-sized $B$ to the complementary subsystem $S^c$, and $S^c$ will generally involve at least half the degrees of freedom in the whole system. To deal with this large $S^c$, we use the Lieb-Robinson bound to show that a mean-field approximation to the time evolution remains reasonably good for sufficiently short times, provided the initial state has product form. For as long as the mean-field approximation holds, the dynamics  cannot generate any significant entanglement, which prohibits signalling to $S^c$ and, of course, scrambling. 

\subsection{Scrambling implies signalling} \label{subsec:scrambling.signal}

Any information initially stored as a state on $\cH_1$ will have become inaccessible to measurements on $S$ alone once scrambling has occurred. One way of phrasing this mathematically is by introducing a ``reference'' system $N$ that does not participate in the interaction and will initially be entangled with system $\cH_1$. The scrambling condition ensures that 
if the initial state has the form $\ket{\Psi(0)} = \ket{\ps_1}_{N\cH_1} \ox \ket{\ps_2}_{\cH_2} \ox  \cdots \ket{\ps_n}_{\cH_n}$, then the time evolution destroys any entanglement between $N$ and $\cH_1$ in the sense that 
\begin{equation} \label{eqn:decoupling}
\left\| \Psi_{N S}(t_*) - \Psi_N(0) \ox \Psi_S(t_*) \right\|_1 \leq \e \operatorname{rank} \Psi_N(0).
\end{equation}
(See, \emph{e.g.}, Lemma 19 of \cite{hayden:2010a}.) To study signalling of a single bit's worth of information, it suffices to let $\ket{\psi_1}_{N \cH_1} = \smfrac{1}{\sqrt{2}} ( \ket{0}_N \ket{0}_{\cH_1} + \ket{1}_N\ket{1}_{\cH_1})$ for a some orthonormal states $\ket{0}$ and $\ket{1}$.

\begin{figure}
\begin{center}
\includegraphics[height=6cm]{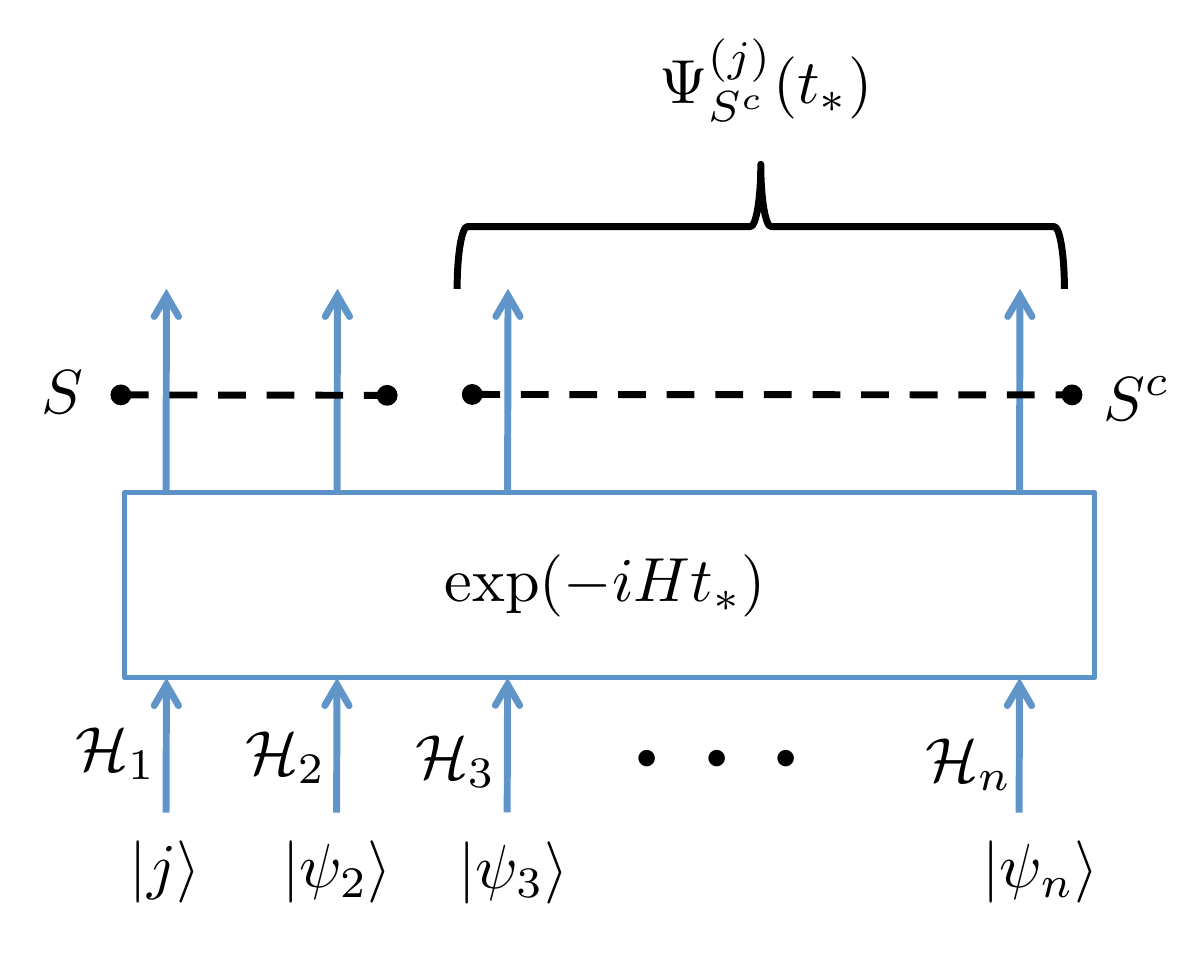}
\caption{Scrambling implies signalling. Site $1$ is prepared in one of two orthogonal states $\ket{j}$ for $j$ either $0$ or $1$. All other sites are prepared in states that are independent of $j$. After the scrambling time $t_*$ any subsystem $S$ of size at most $\kappa n$ will be essentially independent of $j$, but the reduced states $\Psi_{S^c}^{(j)}(t_*)$ on the complementary subsystem $S^c$ will be nearly orthogonal. Scrambling therefore implies signalling from the first site to the complementary system $S^c$.} \label{fig:signal1}
\end{center}
\end{figure}

 As discussed in \cite{hayden:2007a,hayden:2008a}, inequality (\ref{eqn:decoupling}) implies that the entanglement with $N$ can be recovered without use of the degrees of freedom of $S$. That means there is a unitary transformation $V$ on $S^c$ and a qubit subsystem $M$ of $S^c$ such that 
\begin{equation} \label{eqn:decoder}
{}_{NM} \bra{\Phi} \tr_{S^c \backslash M} \left[ (I_N \ox V) \Psi_{NS^c}(t) (I_N \ox V^\dg) \right] \ket{\Phi}_{NM} \geq 1 - 2 \e
\end{equation}
for the maximally entangled state $\ket{\Phi} = \smfrac{1}{\sqrt{2}} ( \ket{00} + \ket{11} )$. The ability to send entanglement to $S^c$ in this way is at least as strong as mere signalling, however. Working from (\ref{eqn:decoder}), standard manipulations imply that if $\cH_1$ were prepared in one of two orthogonal initial states $\ket{0}_{\cH_1}$ and $\ket{1}_{\cH_1}$ then there are orthogonal projectors $\Pi_0$ and $\Pi_1$ on $S^c$ such that
\begin{equation}
\frac{1}{2} \tr \Pi_0 \Psi_{S^c}^{(0)}(t_*)  + 
\frac{1}{2} \tr \Pi_1 \Psi_{S^c}^{(1)}(t_*)
\geq 1 - 4 \e,
\end{equation}
where $\ket{\Psi^{(j)}(0)} = \ket{j}_{\cH_1} \ox \ket{\ps_2}_{\cH_2} \ox \cdots \ox \ket{\ps_n}_{\cH_n}$. That is, the signal has been transmitted from $\cH_1$ to $S^c$ with an average probability of error in the decoding of at most $4\e$. These conclusions are illustrated in Figure~\ref{fig:signal1}.

\subsection{Lieb-Robinson bounds for nonlocal interactions}\label{LR}

As has been the case throughout the paper, the state space will have the form $\cH = \cH_1 \ox \cdots \ox \cH_n$.
Suppose that the Hamiltonian has the two-body form $H = \sum_{\pair{x}{y}} H_{\pair{x}{y}}$,
 where the sum is over unordered pairs of sites $\pair{x}{y}$ and each of $x$ and $y$ 
range from $1$ to $n$. Each term  $H_\pair{x}{y}$ acts only on $\cH_x \ox \cH_y$. We can associate to such a Hamiltonian an {\it interaction graph} $G=(V,E)$ with $n$ vertices 
representing Hilbert spaces $\cH_1,\ldots,\cH_n$ and edges connecting vertices $x$ and $y$ if the 
2-body interaction term $H_{\pair{x}{y}}$ is nonzero. The antiferromagnetic Ising interactions discussed in Section~\ref{sec:ising} are a special case, and the graph of Figure~\ref{fig:graph} is, of course, the interaction graph.
Denote by $D$ the maximum degree of any vertex in the interaction graph. 
Let us further require the constraint $\|H_\pair{x}{y} \| \leq c/D$ on the strength 
of pairwise interactions for some constant $c$. Physically, this constraint ensures that the energy per degree of freedom will remain finite  for all states even in the limit $n\rar \infty$.

For $X \subseteq \{ 1, 2, \ldots, n \}$, denote by $\cA_X$ the algebra of bounded norm operators acting on $\cH_X$. We start by
 discretizing time into steps of size $\epsilon=t/N$ for some large integer $N$  and let 
$t_j=j\epsilon$.
Then, for observables $O_A \in \cA_A$ and $O_B \in \cA_B$,
\begin{equation} \label{commut}
[O_A(t),O_B]=[O_A,O_B]+\sum_{j=0}^{N-1} 
\left([O_A(t_{j+1}),O_B]-[O_A(t_j),O_B]\right).
\end{equation}
The observable $O_A$ evolves after time $\epsilon$ to $O_A(\epsilon)=e^{ih \epsilon} \: O_A\: e^{-i h \epsilon}+O(\epsilon^2)$, 
with $$h=\sum_{x \in A} \sum_{z} H_{\pair{x}{z}}.$$
The norm of each term of the sum in (\ref{commut}) can be bounded from above using
\begin{multline}
\|[O_A(t_{j+1}),O_B]\|=\|[e^{ih\epsilon}O_Ae^{-ih\epsilon},O_B(-t_j)]\|+O(\epsilon^2) \\
	\leq \|[O_A,O_B(-t_j)]\|
+\epsilon\|[O_A,[h,O_B(-t_j)]]\|+O(\epsilon^2).\nonumber 
\end{multline}
Hence, we have
\bea
\|[O_A(t),O_B]\|\leq \|[O_A,O_B]\|+2\epsilon\|O_A\|\sum_{j=0}^{N-1}\|[h,O_B(-t_j)]\|+O(\epsilon). 
\eea
 In the limit $\epsilon\to 0$, the above expression becomes the inequality
\bea\label{ineq}
\|[O_A(t),O_B]\|\leq \|[O_A,O_B]\|+2\|O_A\|\sum_{x \in A} \sum_z\int_0^t ds\|[H_{\pair{x}{z}}(s),O_B]\|.
\eea
We now specialize to the case where $B$ is the singleton set $\{ y \}$. Fixing attention on a particular $O_B \in \cA_y$, define
\begin{equation}\label{defC_B}
C_B(X,t) = \sup_{O_X \in \cA_X} \frac{\| [ O_X(t), O_B] \|}{\| O_X \|}.
\end{equation}
If the subsystem $X$ in the inequality (\ref{ineq}) is $A$, we have
\bea\label{REbound}
C_B(A,t)\leq C_B(A,0)+2\sum_{x\in A}\sum_z\int_0^t ds\: C_B(\pair{x}{z},s)\|H_{\pair{x}{z}}\|,
\eea
whereas for $X=\pair{x}{z}$ we obtain
\bea
C_B(\pair{x}{z},t)\leq C_B(\pair{x}{z},0)+2\sum_{\substack{z_1,z'_1:\\\pair{x}{z}\cap\pair{z_1}{z'_1}\neq 
\emptyset}} \int_0^t ds\: C_B(\pair{z_1}{z'_1},s) \|H_\pair{z_1}{z'_1}\|.
\eea
By using the above bound iteratively in (\ref{REbound}), we find
\begin{multline}\label{rep}
C_B(A,t)\leq C_B(A,0)+2\sum_{x\in A}\sum_z\int_0^t ds\:C_B(\pair{x}{z},0)\|H_{\pair{x}{z}}\| \\
+4\sum_{x\in A}\sum_{\substack{z,z_1,z'_1:\\\pair{x}{z}\cap\pair{z_1}{z'_1}\neq \emptyset}}
\int_0^t ds\int_0^s\:d\tilde{s}\:C_B(\pair{z_1}{z'_1},0)\|H_{\pair{z_1}{z'_1}}\| \|H_{\pair{x}{z}}\|+...
\end{multline}
%
%
By definition, at time $t=0$, the function $C_B(\pair{z}{z'},0)$ is zero unless $z = y$ or $z'=y$. Moreover, from the definition in (\ref{defC_B}),
 it is clear that $C_B(\pair{z}{y},0)\leq 2\|O_B\|$. Thus,
\begin{multline}\label{eqn:lr_general}
C_B(A,t)\leq (2t)\:2|A|\|O_B\|\|H_{\pair{x}{z}}\|+\frac{(2t)^2}{2}\:4|A|\|O_B\|\sum_{\substack{z,z_1:
\\\pair{x}{z}\cap\pair{z_1}{y}\neq \emptyset}}\|H_{\pair{z_1}{y}}\| \|H_{\pair{x}{z}}\| \\
+\frac{(2t)^3}{3!}\:4|A|\|O_B\|\sum_{\substack{z,z_1,z'_1,z_2:\\\pair{x}{z}\cap\pair{z_1}{z'_1}\neq 
\emptyset\\\pair{z_1}{z'_1}\cap\pair{z_2}{y}\neq\emptyset}}\|H_{\pair{z_2}{y}}\|\|H_{\pair{z_1}{z'_1}}\| \|H_{\pair{x}{z}}\|+...
\end{multline}

\begin{figure}
\begin{center}
\includegraphics[height=3cm]{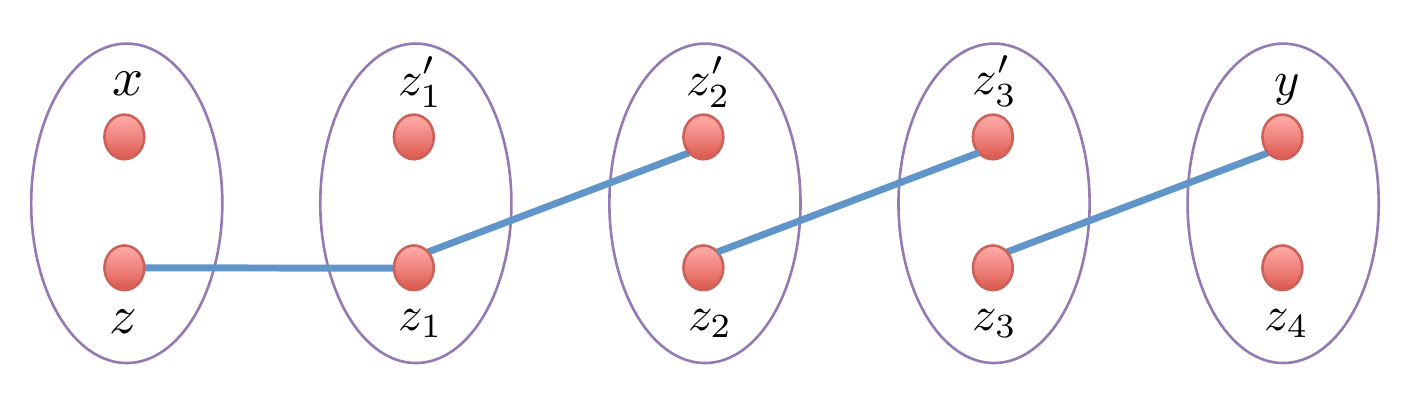}
\end{center}
\caption{Proving the Lieb-Robinson bound on a graph involves a sum over pairs of vertices that contain paths between $x$ and $y$. Starting with 
a set of edges, paths can be visualized for the purpose of counting as different ways of identifying vertices in successive edges of a sequence.
For example, in the figure, each bubble represents an edge and the 
blue lines indicate the identified vertices: $z = z_1 = z_2'$, $z_2=z_3'$ and $z_3 = y$.
There is therefore a path with the following edges: $\pair{x}{z}$, $\pair{z}{z'_1}$, $\pair{z}{z_2}$, $\pair{z_2}{y}$ and $\pair{y}{z_4}$.}
 \label{fig:2.body.counting}
\end{figure}
On a graph of maximum vertex degree $D$, the $i^{th}$ sum in the right hand side of 
(\ref{eqn:lr_general}) has at most $4(4D)^{i-1}$ terms, which can be seen by a simple combinatorial argument. 
The sums that appear have the form
\bea\label{sumcomb}
\sum_{\substack{z,z_1,z'_1,...,z_{i-1}:\\\pair{x}{z}\cap\pair{z_1}{z'_1}\neq 
\emptyset\\\vdots\\\pair{z_{i-2}}{z'_{i-2}}\cap\pair{z_{i-1}}{y}\neq\emptyset}}\left(\frac{c}{D}\right)^i.
\eea
One can think of terms in the above sum as paths made from edges that connect $y$ and $x\in A$, as illustrated in Figure \ref{fig:2.body.counting}.  
A path is made by identifying a vertex in each pair $\pair{z_j}{z'_j}$ with a vertex in $\pair{z_{j+1}}{z'_{j+1}}$. 
Once a vertex $z_j$ is identified with some $z_{j+1}$, there is a maximum of $D$ different choices for $z'_{j+1}$
 because the interaction graph has maximum degree $D$. The path starts either at $x$ or $z$ and ends either at $z_{i-1}$ or $y$. 
For each of these cases, it is not hard to see that the number of paths of length $i$ is less than $(4D)^{i-1}$. 
Therefore, the overall number of terms in the sum (\ref{sumcomb}) is always less than $4(4D)^i$.

Moreover, from the constraint $\|H_\pair{z}{z'} \| \leq c/D$ on the strength of two-body 
interactions, it follows that each term is bounded above by 
$(c/D)^i$. Therefore,
\begin{eqnarray}\label{eqn:lr_bound}
 C_B(A,t)\leq 4|A|\|O_B\|\sum_{i=1}^\infty \frac{(2t)^i}{i!}\left(\frac{c}{D}\right)^i4(4D)^{i-1}<\frac{4|A|\| O_B\|}{D}\:e^{8ct}.
\end{eqnarray}
Finally, note that the arguments of this section are not restricted to the two-body case. Appendix~\ref{sec:hypergraph} shows, for example, that a very similar bound holds for Hamiltonians structured like that of the BFSS matrix model.

If it were possible to signal from $B$ to $A$ in time $t_{signal}$, then there would exist unit norm operators $O_A \in \cA_A$ and $O_B \in \cA_B$ such that 
$\bra{\Psi(0)} [ O_A(t), O_B ] \ket{\Psi(0)} > \d$
for some $\d = O(1)$.
A direct application of (\ref{eqn:lr_bound}) then implies that
\begin{equation}
\label{bbound}
	t_{signal} > \frac{1}{8 c}\log\left(\frac{D\delta}{4|A|}\right).
\end{equation}
In the case of a fully connected graph, $D = n-1$, which would seem to force logarithmic scaling of the signalling and, therefore, of the scrambling time. Unfortunately, as discussed in Section~\ref{subsec:scrambling.signal}, scrambling only implies signalling to $S^c$ so we must take $A = S^c$, and systems of size larger than $n/2$ don't scramble, so $|S^c| \geq n/2$. Na\"ive substitution into (\ref{bbound}) then yields no bound at all on the scrambling time so further analysis will be necessary.

\subsection{Scrambling highly mixed initial states} 

\begin{figure}[t]
\begin{center}
\includegraphics[height=7.5cm]{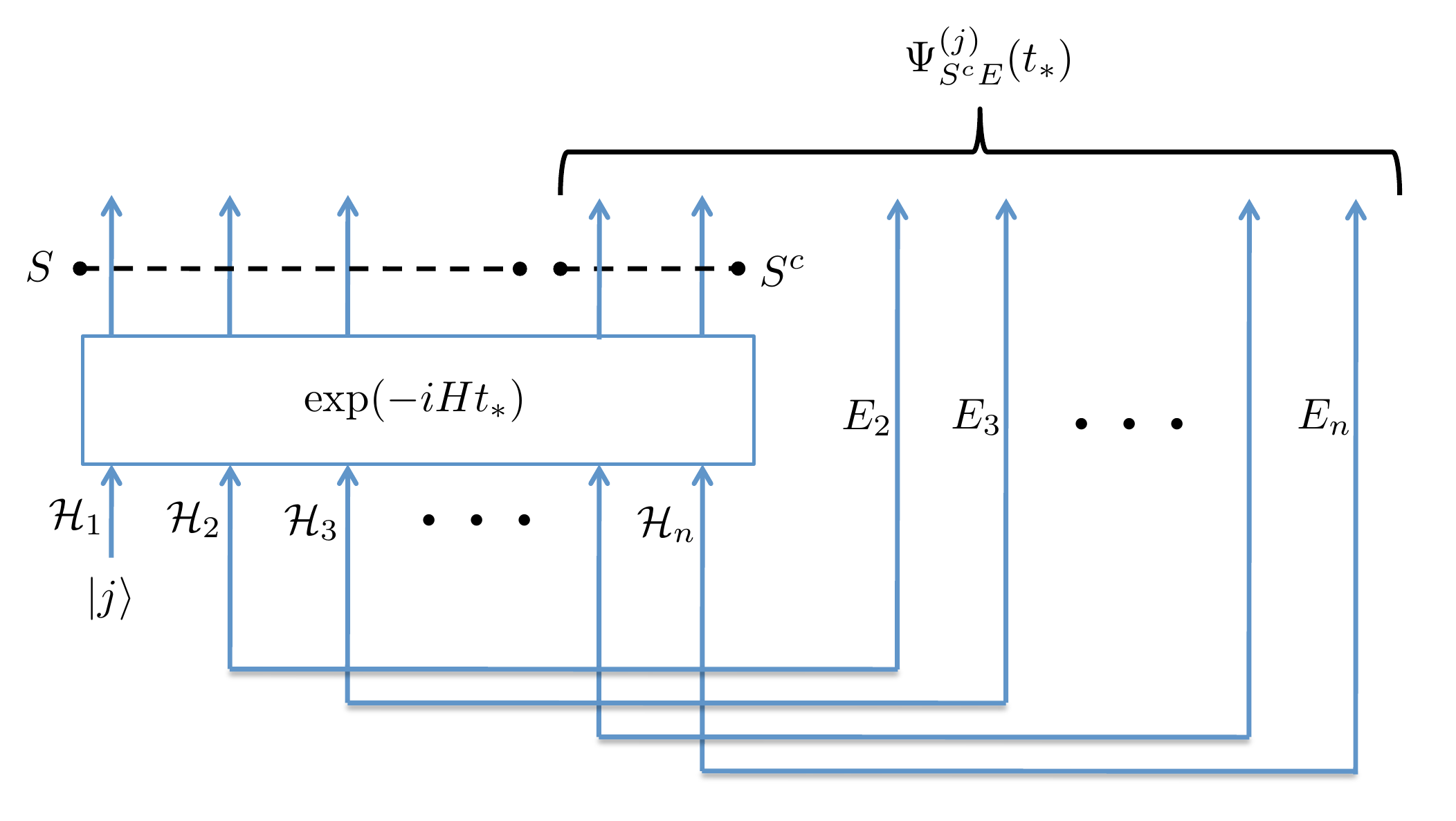}
\caption{Scrambling implies signalling for mixed initial states. Site $1$ is prepared in one of two orthogonal states $\ket{j}$ for $j$ either $0$ or $1$, and all other states are prepared in states that are independent of $j$ and highly mixed. These mixed states can be viewed as parts of pure states that are entangled with environmental degrees of freedom $E_2$ through $E_n$. When the initial states are \emph{maximally} mixed, it is possible to scramble subsystems $S$ of size $n - O(1)$. This leads to  signalling  to the complementary degrees of freedom $S^c$, adjoined with the environmental degrees of freedom $E = E_2 \cdots E_n$. That is, the states $\Psi_{S^c E}^{(j)}(t_*)$ are nearly orthogonal to each other. Because $S^c$ can be taken to be constant-sized, the Lieb-Robinson bound provides nontrivial lower bounds on the signalling, and hence scrambling, time in this setting without the need for additional argument.} \label{fig:signal2}
\end{center}
\end{figure}

It's interesting to note that (\ref{bbound}) \emph{does} yield a logarithmic lower bound for the type of scrambling relevant to information retrieval from highly entangled black holes. This paper has thus far focused exclusively on pure initial states for $\cH$. Replacing $\ket{\Psi(0)}$ with a state pure on $\cH_1$ and maximally mixed on $\cH_2$ through $\cH_n$ corresponds to a different communication scenario. The retrieval of the information stored in $\cH_1$ would need to make use of some degrees of freedom $S^c \subseteq \{ 1, 2, \ldots n \}$ supplemented by the environmental degrees of freedom required to ``purify'' the initial state. When the initial state is so highly mixed, however, it is possible to scramble many more degrees of freedom than when the initial state is pure, leading to a much smaller $S^c$. The resulting signalling scenario is illustrated in Figure~\ref{fig:signal2}. Brownian quantum circuits, for example, will scramble subsystems $S$ of size $n - O(1)$, leaving a constant-sized complementary system $S^c$ with $|S^c| = O(1)$. Because the environmental degrees of freedom don't participate in the interaction, one can take $|A| = |S^c| = O(1)$ and recover the logarithmic lower bound on scrambling from (\ref{bbound}). Moreover, it is \emph{necessary} to consider these larger systems: numerical investigations show that it is possible to scramble any constant fraction of the degrees of freedom in constant time if the initial state is highly mixed.

\subsection{Controlled mean-field approximation via Lieb-Robinson} \label{subsec:mean.field}

Having proven the Lieb-Robinson bound, we now prove that up to times of order $\log(D)$, the reduced density matrix on each site $x$ is close to a pure state. Since scrambling requires entanglement, this will provide the desired lower bound on the scrambling time.
Since $D$ is the maximum vertex degree, this evaluates to an order $\log(n)$ lower bound for Hamiltonians in which every degree of freedom interacts with a constant fraction of all the others.

A slightly subtle point is that all of a system's single site density operators can in principle be close to pure even if the wavefunction of the whole system is not. The issue is that the number of sites, $n$, is large, and the overlap of the true wavefunction with the mean-field pure product state can easily be a factor exponentially smaller in $n$ than the corresponding single-site overlap.
The analysis of this section will therefore \emph{not} imply that the wavefunction of the whole system is product up to times of order $\log(D)$.

We begin by defining a time-dependent ``mean-field" Hamiltonian
\be
H^{MF}(t)=\sum_x H^{MF}_x(t),
\ee
where each operator $H^{MF}_x$ is supported on site $x$.
We define the operators $H^{MF}_x$ self-consistently as follows.  Let $\Psi^{MF}_x(t)$ be the reduced density matrix on site $x$ at time $t$ assuming that the state is initialized to a product state $\ket{\Psi(0)} = \ket{\ps_1}_{\cH_1} \ox \cdots \ox \ket{\ps_n}_{\cH_n}$ at time $t=0$ and evolves under Hamiltonian $H^{MF}$.  Then
\be
\partial_t \Psi^{MF}_x(t)=-i[H^{MF}_x(t),\Psi^{MF}_x(t)].
\ee
We then define
\be
H^{MF}_x(t)=\sum_y {\rm tr}_y \Bigl(H_{\langle x,y \rangle} \Psi^{MF}_y(t)\Bigr).
\ee
This provides the self-consistent definition of $H^{MF}$.
We also define
\be
H^{MF}_{\langle x,y \rangle}={\rm tr}_y \Bigl(H_{\langle x,y \rangle} \Psi^{MF}_y(t)\Bigr),
\ee
so that $H^{MF}_x=\sum_{y} H^{MF}_{\langle x,y\rangle}$.

Define $\Psi_x(t)$ to be the reduced density matrix of the state evolving under Hamiltonian $H=\sum_{\langle x,y \rangle} H_{\langle x,y \rangle}$ again assuming that the state is initialized to the product state $\ket{\Psi(0)}$ at time $t=0$.  We now prove that, for $t$ small compared to $\log(n)$,  $\Psi_x(t)$ is close (in trace norm distance) to $\Psi^{MF}_x(t)$.

For notational convenience, we will write $\langle O \rangle$ to indicate $\langle \Psi(0)  | O | \Psi(0) \rangle$.  Further, we define
a unitary $U^{MF}_x(t)$ to define the mean-field evolution on site $x$ by
\be
U^{MF}_x(0)=I,
\ee
and
\be
\partial_t U^{MF}_x(t)=-i H^{MF}_x(t) U^{MF}_x(t).
\ee
Also, define
\be
U^{MF}_x(t,s)=U^{MF}_x(t) U^{MF}_x(s)^\dagger.
\ee
That is, $U^{MF}_x(t,s)$ describes mean-field evolution from times $s$ to time $t$.
Since the mean-field time evolution on all of $\cH$ for time $t$ has the form $\ox_{x=1}^n U_x^{MF}(t)$, it can never generate any entanglement between different sites. For as long as it remains a decent approximation to the true time evolution, scrambling will be impossible.

Similarly, we define $U(t)$ to be the unitary describing evolution under $H$, with
\be
U(0)=I,
\ee
and
\be
\partial_t U(t)=-i H U(t).
\ee
Define $U(s,t)=U(s) U(t)^\dagger$.

In proving the Lieb-Robinson bound above in Section~\ref{LR},
we used the Heisenberg notation for operator evolution:  $O(t)$ denoted $U(t)^\dg O U(t)$.
In this section, we will not use this Heisenberg notation, and we will instead explicitly write out $U(t)$ or $U(t)^\dagger$ to describe evolution of operators or states.  The reason for this is that we 
are going to evaluate the expectation values of operators whose time-dependence is not necessarily given by conjugation by $U(t)$, so that the parenthetical $(t)$ could be ambiguous if we were to use it to denote Heisenberg evolution.

Consider any operator $O_x$ supported on site $x$.
For any two times, $t_i$ and $t_f$, we have
\begin{multline}
U(t_f,t_i)^\dagger  O_x U(t_f,t_i)
= U^{MF}_x(t_f,t_i)^\dagger O_x U^{MF}_x(t_f,t_i) \\  +i \int_{t_i}^{t_f} {\rm d}s  \,
U(s,t_i)^\dagger \Bigl[\Bigl( H-H^{MF}_x(s) \Bigr),  U^{MF}_x(t_f,s)^\dagger O_x U^{MF}_x(t_f,s)\Bigr] U(s,t_i).
\end{multline}
This equation can be proven by differentiating the right-hand side with respect to $t_i$ and verifying that the result is equal to the right-hand side multiplied by $i$ and commuted with $H$.  Call the first and second terms on the right-hand side $T_1$ and $T_2$ respectively.  When $T_1$ is differentiated with respect to $t_i$ the result is $i[H^{MF}_x(t_i),T_1]$, while differentiating $T_2$ with respect to $t_i$ gives two terms, one from the change in the limit of the integral (this term is equal to i$[H-H^{MF}_x(t_i),T_1]$ and adding this to the derivative of $T_1$ respect to $t_i$ gives $i[T_1,H]$) and one term from the change in $U(s,t_i)$ which gives $i[T_2,H]$.
Specializing to $t_i=0$,
we have
\begin{multline}
\label{withcom}
U(t)^\dagger  O_x U(t)
=U^{MF}_x(t)^\dagger O_x U^{MF}_x(t) \\  +i \int_0^{t} {\rm d}s \, 
U(s)^\dagger \Bigl[\Bigl( H-H^{MF}_x(s) \Bigr), U^{MF}_x(t,s)^\dagger O_x U^{MF}_x(t,s)\Bigr] U(s).
\end{multline}
We will apply this equation to the specific case of the time-dependent operator $O_x=1-\Psi_x^{MF}(t)$, using it to compute the expectation value
\begin{equation}
\langle U(t)^\dagger (1-\Psi_x^{MF}(t) ) U(t) \rangle
 = 1 - \bra{\Psi_x^{MF}(t)} \Psi_x(t) \ket{\Psi_x^{MF}(t)}
\end{equation}
and show that  $\Psi_x(t)$ is close to $\Psi^{MF}_x(t)$.

Note that for any operator $O$ supported on $x$, we have
\be
\label{justcom}
\Big[ H-H^{MF}_x(s),O\Big]=\sum_y \Bigl[ H_{\langle x,y \rangle}-H^{MF}_{\langle x,y \rangle},O \Big].
\ee
This holds in particular
in the case that $O=U_x^{MF}(t,s)^\dagger O_x U_x^{MF}(t,s)$ as in Eq.~(\ref{withcom}).
 For any given $y$, the trace
${\rm tr}_y (\Psi^{MF}_y(s) ( H_{\langle x,y \rangle}-H^{MF}_{\langle x,y \rangle}))=0$.
Note also that $\Psi^{MF}_y(s)$ is a projector for all $y$ and $s$. Write
\be
\label{defby}
H_{\langle x,y \rangle}-H^{MF}_{\langle x,y \rangle}(s)=L_{\langle x,y \rangle}(s)+R_{\langle x,y \rangle}(s),
\ee
where
\be
L_{\langle x,y \rangle}(s)=(1-\Psi_y^{MF}(s)) \Bigl( H_{\langle x,y \rangle}-H^{MF}_{\langle x,y \rangle}(s) \Bigr),
\ee
and $R_{\langle x,y \rangle}(s)$ is defined by Eq.~(\ref{defby}).  Then,
$\Psi_y^{MF}(s) L_{\langle x,y \rangle}(s)=0$ and similarly $R_{\langle x,y \rangle}(s) \Psi_y^{MF}(s)=0$.

Taking into account Eq.~(\ref{justcom}) as well as the definitions of $L$ and $R$, we can replace $H-H_x^{MF}(s)$ in Eq.~(\ref{withcom}) with a sum over $y$ of
$L_{\langle x,y \rangle}(s)+R_{\langle x,y \rangle}(s)$.
This gives a sum over $y$ of a sum of two terms (the $L$ and $R$ terms).
Consider an $L$ term for given $y,s$.  This is
\begin{eqnarray}
\label{eqL}
&&
U(s)^\dagger [ L_{\langle x,y \rangle}(s),U^{MF}_x(t,s)^\dagger O_x U^{MF}_x(t,s)] U(s)
\\ \nonumber
&=&
U(s)^\dagger (1-\Psi_y^{MF}(s)) [ L_{\langle x,y \rangle}(s),U^{MF}_x(t,s)^\dagger O_x U^{MF}_x(t,s)] U(s)
\\ \nonumber
&=&
\Bigl( U(s)^\dagger (1-\Psi_y^{MF}(s)) U(s) \Bigr) 
\\ \nonumber
&& \times\Bigl( U(s)^\dagger [ L_{\langle x,y \rangle}(s),U^{MF}_x(t,s)^\dagger O_x U^{MF}_x(t,s)] U(s)\Bigr).
\end{eqnarray}
Similarly, for an $R$ term, we write
\begin{eqnarray}
\label{eqR}
&&
U(s)^\dagger [ R_{\langle x,y \rangle}(s),U^{MF}_x(t,s)^\dagger O_x U^{MF}_x(t,s)] U(s)
\\ \nonumber
&=&
U(s)^\dagger [ L_{\langle x,y \rangle}(s),U^{MF}_x(t,s)^\dagger O_x U^{MF}_x(t,s)]  (1-\Psi_y^{MF}(s)) U(s)
\\ \nonumber
&=&
\Bigl( U(s)^\dagger [ L_{\langle x,y \rangle}(s),U^{MF}_x(t,s)^\dagger O_x U^{MF}_x(t,s)] U(s)\Bigr)
\\ \nonumber
&& \times \Bigl( U(s)^\dagger (1-\Psi_y^{MF}(s))U(s)\Bigr).
\end{eqnarray}
For an $L$ term, we apply Eq.~(\ref{withcom}) to the first term
$\Bigl( U(s)^\dagger (1-\Psi_y^{MF}(s)) U(s) \Bigr)$ on the last line of Eq.~(\ref{eqL}), while for an $R$ term, we apply Eq.~(\ref{withcom}) to the last term $\Bigl( U(s)^\dagger (1-\Psi_y^{MF}(s))U(s)\Bigr)$ on the last line of Eq.~(\ref{eqR}).

We proceed iteratively in this fashion, getting an infinite series of terms.  Each term in the series at a given order, say the $k$-th order, involves a $k$-fold integral over $s_1,s_2,\ldots,s_k$, with $0\leq s_1 \leq \cdots\leq s_k \leq t$.  Further, each term in the series has a sum over $k$ different sites $y_1,y_2,\ldots,y_k$ and finally each term has a sum over $k$ different choices of $L$ or $R$ terms.
Our goal is to bound the expectation of the sum of terms at $k$-th order.  Each such term will have one operator $1-\Psi_{y_k}(s_k)$ in it.  This operator may be in the middle of a sequence of terms.  Suppose the last term was an $L$ term.  Then we have some expectation value
\be
\label{PQexval}
\left\langle P \Bigl( U^{MF}(s)^\dagger (1-\Psi_{y_k}^{MF}(s)) U^{MF}(s) \Bigr) Q \right\rangle
\ee
for some operators $P,Q$.  We commute $\Bigl( U^{MF}(s)^\dagger (1-\Psi_{y_k}^{MF}(s)) U^{MF}(s) \Bigr)$ through $P$ using the Lieb-Robinson bounds above.  Note that the reason that we choose to commute through $P$ rather than through $Q$ is that whenever the last term is an $L$ term, one of the operators in $Q$ is $L_{\langle y_{k-1},y_k \rangle}$. We would not be able to bound the associated commutator since $Q$ has support on $y_k$.  Conversely, if the last term was an $R$ term, we commute to the right through $Q$ instead.
Note that 
\begin{equation}
\left\langle \Bigl( U^{MF}(s)^\dagger (1-\Psi_{y_k}^{MF}(s)) U^{MF}(s) \Bigr) S \right\rangle
=\left\langle S \Bigl( U^{MF}(s)^\dagger (1-\Psi_{y_k}^{MF}(s)) U^{MF}(s) \Bigr) \right\rangle
=0
\end{equation}
 for any operator $S$.  Therefore, the expectation value Eq.~(\ref{PQexval}) is bounded by the commutator
$\Vert [P,\Bigl( U^{MF}(s)^\dagger (1-\Psi_{y_k}^{MF}(s)) U^{MF}(s) \Bigr)] \Vert$ in the case that the last term was an $L$ term. (Similarly. it is bounded by a commutator with $Q$ in the case of an $R$ term.)

To bound this commutator, we consider two different cases. 
First, there is the case that $y_k \neq y_i$ for $1 \leq i <k$.  
In this case, we can bound the
commutator
by
$({\rm const.}/D)^k \times (k/D) \times \exp({\rm const.} \times t)$ using the Lieb-Robinson bound from Section~\ref{LR}, which contributes a factor of ${\rm const.} \times (1/D) \times \exp({\rm const.} \times t)$. The factor of $k$ appears because $P$ is a product of up to $k$ different operators while the final factor of $({\rm const.}/D)^k$ comes from the fact the norms of all of the operators $L_{\langle x,y \rangle}$ and $R_{\langle x,y \rangle}$ are bounded above by ${\rm const} / D$.
The case when $y_k=y_i$ for some $1 \leq i <k$ might seem to be more problematic because the Lieb-Robinson bound doesn't apply, but we will see below that this bad case happens infrequently enough to not affect the final conclusion.

To bound the sum over terms in the series at given order, we note that the sum over choices of $y_1,\ldots,y_k$ decomposes into these same two cases. The sum in the first case, when $y_k \neq y_i$ for all $1 \leq i < k$, is bounded by
\be
\label{case1}
{\rm const.} \times \left(\frac{k}{D}\right) \frac{({\rm const.} \times t)^k}{k!} \exp({\rm const.}\times t),
\ee
 where the factor of $(k/D) \exp({\rm const.}\times t)$ is due to the commutator bound, with the factor of $1/D^k$ that was present there cancelled by an $D^k$ in the numerator arising from the sum over $y_1,\ldots,y_k$.
The factor of $t^k/k!$ in Eq.~(\ref{case1}) arises from integrating over the $k$ different times $0\leq s_1 \leq \cdots\leq s_k \leq t$. Summing over the different choices of $L$ or $R$ contributes an extra factor of $2^k$ which can be absorbed into the constant raised to the power $k$.
In the second case, when $y_k = y_i$ for at least one $1 \leq i < k$, the sum over $y_i$ is bounded by
${\rm const.} \times (k/D) \times \frac{({\rm const.} \times t)^k}{k!}$ where the factor of $k/D$ arises because any of the $k-1$ different $y_i$ for $1\leq i <k$ may be equal to $y_k$. (By constraining the choice of $y_i$ we reduce the number of different choices for $y_i$ in the sum.)

So, the sum over all orders $k$ is bounded by  
\begin{eqnarray}
\nonumber
&& {\rm const.} \times \sum\limits_{k=1}^{\infty}\left(\frac{k}{D}\right) \times \frac{({\rm const.} \times t)^k}{k!} \times \exp({\rm const.} t)\\
& \leq &{\rm const.} \times (1/D) \times \exp({\rm const.} \times t). \label{eqn:enchilada}
\end{eqnarray}

Recall that this is an upper bound on the quantity
$1 - \bra{\Psi_x^{MF}(t)} \Psi_x(t) \ket{\Psi_x^{MF}(t)}$, the deviation of $\Psi_x(t)$ from being a pure state. If the deviation is small at time $t$, the continuity of the von Neumann entropy implies that $H(\Psi_x(t)) \leq \d \log \dim \cH_x$ for some universal $\d$ going to zero with the deviation~\cite{fannes:73a}. The subadditivity property of $H$ then implies that
\begin{equation}
H(\Psi_S(t)) \leq \sum_{x \in S} H( \Psi_x(t)) \leq \d  \log \dim \cH_S.
\end{equation}
As discussed in Section \ref{subsec:scrambling.entanglement}, scrambling requires that $H(\Psi_S(t))$ be close to its maximal value of $\log \dim \cH_S$, which can only occur if the deviation of each $\Psi_x(t)$ is significant. For this to happen, (\ref{eqn:enchilada}) requires that $t$ be order $\log(D)$, which is the desired lower bound on the scrambling time provided $D \sim n$. (Note that it is equally possible, if slightly more technical, to supply a dimension-independent argument.)

\subsection{Sparse graphs} \label{subsec:sparse}

If the degree $D$ is constant or even scaling sublinearly with $n$, then (\ref{eqn:enchilada}) might not be a useful bound. For sufficiently slowly growing $D$, however, it is possible to substitute the more traditional Lieb-Robinson bound for the version proved in Section~\ref{LR}. Specifically, the version of the bound proved in~\cite{hastings:2006a} ensures that 
\begin{equation} \label{eqn:LR.usual}
\big\| [ O_A(t), O_B ] \big\| \leq {\rm const.} \times \exp\left[ (vt - d(A,B))/\xi \right] \| O_A \| \| O_B \|
\end{equation}
for some positive constants $v$ and $\xi$. The function $d(A,B)$ measures the distance from $A$ to $B$ in the interaction graph so the interpretation of (\ref{eqn:LR.usual}) is that there is a maximum effective velocity $v$ of information propagation between degrees of freedom. For complete graphs, the bound is trivial, but not for graphs of lower connectivity.

In particular, there can be at most $D^l$ vertices at distance exactly $l$ from any fixed vertex. 
It follows that at most a fraction $\a$ of all pairs of vertices $x$ and $y$ can satisfy $d(x,y) \leq \log(\a n)/ \log D$. 
Therefore, most $x$ and $y$ satisfy $d(x,y) \geq O(\log n / \log D)$. Substituting into (\ref{eqn:LR.usual}) and comparing with (\ref{eqn:lr_bound}) implies that the signalling time between $x$ and $y$ must satisfy
\begin{equation} \label{eqn:root.bound}
t_{signal} \geq \min \left( O( \log D ), O\left( \frac{\log n}{\log D} \right) \right) \geq O( \sqrt{ \log n } ).
\end{equation}

For regular graphs, in which \emph{every} vertex has degree $D$, this reasoning can even be extended to the scrambling time $t_*$. From the mean-field argument, we already know that $t_* \geq O(\log D)$. A direct application of Lieb-Robinson, however, requires that $t_* \geq O( \log n / \log D )$. To see this, fix $x$ and let $S$ be the set of all sites $y$ such that $d(x,y) \leq \log n/\log(D-1) + {\rm const}$. This will be a constant fraction of all the sites. Different initial states at site $x$ are eigenstates of rank one projectors acting on that site. By a standard argument~\cite{hastings:2004a}, (\ref{eqn:LR.usual}) ensures that for times $t<d(x,S^c)/v - {\rm const.}$, the time-evolved projectors will be well-approximated by operators acting only on $S$, in which case the different initial states can be distinguished by measurements on $S$ alone, which is inconsistent with scrambling. Optimizing over $D$ as in (\ref{eqn:root.bound}) yields $t_* \geq O(\sqrt{\log n})$.

%
%
 
\section{Conclusions}

We have explored two aspects of the fast scrambling conjecture, both of which are implicit 
in the statement that the most rapid scramblers take a time logarithmic in the number of 
degrees of freedom. For the statement to be true,  there must exist systems scrambling quickly
 enough to saturate the bound. Conversely, no system should be capable of scrambling in time 
faster than logarithmic.

We demonstrated that Brownian quantum circuits and the Ising model on sparse random graphs
 both scramble information in logarithmic time. Each example, however, has its own deficiencies,
 not quite meeting the objective of finding a time-independent Hamiltonian that scrambles all 
locally available information in logarithmic time. Namely, Brownian quantum circuits are not 
actually described by a time-independent Hamiltonian, while the Ising model only scrambles 
information in one basis, leaving the conjugate basis invariant. Nonetheless, the examples 
illustrate that the entanglement creation required for scrambling can indeed be accomplished
 in logarithmic time without the need for an intricately structured Hamiltonian.
 Finding a completely fast scrambling time-independent Hamiltonian remains an open problem. While it's simple 
enough to write down plausible candidates, analyzing them is a challenge.

To find limits on scrambling, we used Lieb-Robinson techniques to prove a general lower 
bound on the scrambling time of arbitrary quantum systems with two-body interactions.
 The strategy was to estimate the amount of time required to signal in such systems, which in turn
 bounds the amount of time required to scramble. Mathematically, we used a modified Lieb-Robinson bound to argue that for sufficiently small times, a mean-field approximation to the single-site evolution is a good approximation.
If most pairs of systems interact with terms of comparable norm in the Hamiltonian, the result is a logarithmic 
lower bound on the scrambling time. The same bound applies to four-body Hamiltonians with structure similar to the BFSS matrix model. However, our argument does contain a loophole: 
in the general case of graphs with lower connectivity, we could only prove a requirement that the scrambling time be at least $O(\sqrt{\log n})$, although we strongly suspect that this is only a reflection of the limitations of our technique.


One of the lessons of this investigation is that some plausible mathematical formulations of the
 conjecture are false. In the case of the Ising model, for example, the scrambling time ratio $\tau_* = t_* / t^{(1)}_*$, 
which {\it a priori} one might have thought should also grow at least logarithmically with the number of
 degrees of freedom, is parametrically smaller. More subtly, the fast scrambling conjecture is formulated in terms of pure initial states  and scrambling sets $S$ of size $|S| = \kappa n$ for constant $\kappa$. The argument for rapid release of information from highly entangled black holes, however, requires starting from a mixed initial state and studying larger scrambling sets $S$ of size $n - O(1)$ instead of just $\kappa n$. We have found logarithmic lower bounds on the scrambling time in both cases but not using identical reasoning. The pure state scenario, perhaps surprisingly, was more difficult to analyze.

The understanding gained here should ultimately be helpful in properly formulating and evaluating the 
scrambling time of matrix quantum mechanics or other models of black holes. The correct analog
 of the simple decomposition into subsystems used here already poses a bit of a puzzle. Likewise,
 since some initial configurations are known not to scramble quickly, care is required in 
identifying the set of states that are rendered locally indistinguishable by the dynamics. 
The correct analog of ``local information'' should be physically well-motivated and basis-independent.
 The reward for resolving these issues will be great: a microscopic description of information
 leakage from black holes and, more generally, a deeper understanding of how nonlocal degrees of freedom 
in quantum gravity can be reconciled with the causal nature of semiclassical physics.

\acknowledgments
We are grateful to Ashton Anderson, John Preskill, Norbert Schuch, Luc Vinet and especially both Steve Shenker and Lenny Susskind for helpful conversations. PH would also like to thank the Stanford Institute for Theoretical Physics and the Aspen Center for Physics for their kind hospitality. PH was supported by CIFAR, NSERC, the Canada Research Chairs program and ONR through grant N0001480811249. DS was supported by the United States NSF under the GRF program, as well as grant 0756174.

\begin{appendix}
\section{Equations of motion for Brownian quantum circuits}\label{BrownApp}
In this appendix we describe in detail the dynamics of the purity of the subsystem $S$
as it evolves according to a Brownian quantum circuit. Our starting point is the 
equation of motion for $\Psi_S(t)$.
This can be found by tracing out the degrees of freedom in $S^c$ in (\ref{dpsi}):
\begin{multline}
d\Psi_S(t) = \frac{i}{\sqrt{8n(n-1)}}\sum_{j<k}\sum_{\alpha_j,\alpha_k=0}^3
\tr_{S^c}(\left[\sigma_j^{\alpha_j}\otimes\sigma_k^{\alpha_k}\otimes
 I_{\setminus\{j,k\}}, \Psi(t)\right])
dB_{\alpha_j,\alpha_k}(t) - \Psi_S(t) dt \\
+\frac{1}{8n(n-1)}\sum_{j<k} \sum_{\alpha_j,\alpha_k=0}^3 \tr_{S^c}\left(
\left(\sigma_j^{\alpha_j}\otimes\sigma_k^{\alpha_k}\otimes I_{\setminus\{j,k\}}
\right)\Psi(t)\left(\sigma_j^{\alpha_j}
\otimes\sigma_k^{\alpha_k}\otimes I_{\setminus\{j,k\}}\right)\right)
dt.
\end{multline}
The right hand side of this equation of motion consists of a noisy part
\begin{equation}
(\dagger) = \frac{i}{\sqrt{8n(n-1)}}\sum_{j<k}\sum_{\alpha_j,\alpha_k=0}^3
\tr_{S^c}(\left[\sigma_j^{\alpha_j}\otimes\sigma_k^{\alpha_k}\otimes
 I_{\setminus\{j,k\}}, \Psi(t)\right])
dB_{\alpha_j,\alpha_k}(t)
\end{equation}
and a noiseless part
\begin{equation}
(\dagger\dagger) = - \Psi_S(t) dt\:+
\frac{1}{8n(n-1)}\sum_{j<k} \sum_{\alpha_j,\alpha_k=0}^3 \tr_{S^c}\left(
\left(\sigma_j^{\alpha_j}\otimes\sigma_k^{\alpha_k}\otimes I_{\setminus\{j,k\}}\right)
\Psi(t)\left(\sigma_j^{\alpha_j}\otimes\sigma_k^{\alpha_k}\otimes I_{\setminus\{j,k\}}\right)\right)dt
\end{equation}
We'll deal with both of these terms in turn. First, the noisy part
$(\dagger)$ can be reduced to
\begin{multline}
(\dagger) = \frac{i}{\sqrt{8n(n-1)}}\sum_{j<k\in
S}\sum_{\alpha_j,\alpha_k=0}^3
[\sigma_j^{\alpha_j}\sigma_k^{\alpha_k}, \Psi_S(t)]
dB_{\alpha_j,\alpha_k}(t) \\ + 
\frac{i}{\sqrt{8n(n-1)}}\sum_{\substack{j\in S\\ k\in
S^c}}\sum_{\alpha_j,\alpha_k=0}^3 [\sigma_j^{\alpha_j},
\Psi^{\alpha_k}_S(t)] dB_{\alpha_j,\alpha_k}(t),
\end{multline}
where
\begin{equation}
\Psi_S^{\alpha_k}(t) = \tr_{S^c}(\sigma_k^{\alpha_k}\Psi(t))
\end{equation}
and we have omitted tensor products with the identity to make the expressions more compact.
The noiseless part $(\dagger\dagger)$ can be rewritten as
\begin{equation}
(\dagger\dagger) = - \Psi_S(t) dt + \frac{1}{8n(n-1)} \sum_{j<k}\tr_{S^c}\left(4\:I_{j,k}\otimes\Psi_{\setminus\{j,k\}}\right)
dt,
\end{equation}
which expands to a form that distinguishes different contributions:
\begin{multline}
(\dagger\dagger) = - \Psi_S(t) dt + \frac{1}{2n(n-1)} \sum_{j<k \in S}
\Psi_{S\setminus\{j,k\}}(t)\otimes I_{j,k} \, dt \\
+ \frac{|S^c|}{n(n-1)} \sum_{j\in S} \Psi_{S\setminus
\{j\}}(t)\otimes I_j \, dt +
\frac{|S^c|(|S^c|-1)}{n(n-1)} \Psi_S(t) \, dt.
\end{multline}
Reassembling the pieces yields the final equation of motion for $\Psi_S(t)$:
\begin{multline}
d\Psi_S(t) = \frac{i}{\sqrt{8n(n-1)}}\sum_{j<k\in
S}\sum_{\alpha_j,\alpha_k=0}^3
[\sigma_j^{\alpha_j}\otimes\sigma_k^{\alpha_k}\otimes I_{S\setminus\{j,k\}}, \Psi_S(t)]
\, dB_{\alpha_j,\alpha_k}(t) + \\ 
\frac{i}{\sqrt{8n(n-1)}}\sum_{\substack{j\in S\\ k\in
S^c}}\sum_{\alpha_j,\alpha_k=0}^3 [\sigma_j^{\alpha_j}\otimes I_{S\setminus\{j\}},
\Psi^{\alpha_k}_S(t)] dB_{\alpha_j,\alpha_k}(t) - \Psi_S(t) \, dt + \\
\frac{1}{2n(n-1)} \sum_{j<k\in S}
\Psi_{S\setminus\{j,k\}}(t)\otimes I_{j,k} \, dt  +
\frac{|S^c|}{n(n-1)} \sum_{j\in S} \Psi_{S\setminus
\{j\}}(t)\otimes I_j \, dt + \\
\frac{|S^c|(|S^c|-1)}{n(n-1)}\Psi_S(t) \, dt.
\end{multline}
By another application of Ito's rule, the equation of motion for the purity $h_S(t)$ can be derived from the relation
\begin{equation}\label{dh_S}
dh_S(t) = 2\tr(\Psi_S(t)d\Psi_S(t)) + \tr((d\Psi_S(t))^2).
\end{equation}
Because of the number of terms, it will be necessary to work with the equation of motion in pieces, as we did for $\Psi_S(t)$:
\begin{equation}
dh_S(t) = (*) + (**) + (***),
\end{equation}
where $(*)$ and $(**)$ are, respectively, the noisy and noiseless parts coming from the first term in 
(\ref{dh_S}), and $(***)$ is the contribution of the second term.
Firstly, $(*)$ is given by
\begin{multline}
(*) = \frac{i}{\sqrt{8n(n-1)}}\sum_{j<k\in
S}\sum_{\alpha_j,\alpha_k=0}^3
\tr(\Psi_S(t)[\sigma_j^{\alpha_j}\otimes\sigma_k^{\alpha_k}\otimes I_{S\setminus\{j,k\}}, \Psi_S(t)])
\, dB_{\alpha_j,\alpha_k}(t) + \\
\frac{i}{\sqrt{8n(n-1)}}\sum_{\substack{j\in S\\ k\in
S^c}}\sum_{\alpha_j,\alpha_k=0}^3 \tr(\Psi_S(t)[\sigma_j^{\alpha_j}\otimes I_{S\setminus\{j\}},
\Psi^{\alpha_k}_S(t)]) \, dB_{\alpha_j,\alpha_k}(t).
\end{multline}
There is no need to simplify this term any further because it will average to zero when
we consider $\overline{h}_S$.
The second term is more important for what follows:
\begin{multline}
(**) =  \frac{1}{n(n-1)}\sum_{j\not=k\in S}
h_{S\setminus\{j,k\}}(t)\, dt  + \frac{2|S^c|}{n(n-1)} \sum_{j\in S}
h_{S\setminus \{j\}}(t)\, dt \\ +
 \left(\frac{2|S^c|(|S^c|-1)}{n(n-1)} -2\right)h_S(t) \, dt. 
\end{multline}
Finally, $(***)$ is just $\tr((d\Psi_S(t))^2)$:
\begin{multline}
(***) = \frac{1}{8n(n-1)}\sum_{j<k\in
S}\sum_{\alpha_j,\alpha_k=0}^3
\tr\left([\sigma_j^{\alpha_j}\otimes\sigma_k^{\alpha_k}\otimes I_{S\setminus\{j,k\}}, \Psi_S(t)]^2\right)
dt + \\
\frac{1}{8n(n-1)}\sum_{\substack{j\in S\\ k\in
S^c}}\sum_{\alpha_j,\alpha_k=0}^3 \tr\left([\sigma_j^{\alpha_j}\otimes _I{S\setminus j\}},
\Psi^{\alpha_k}_S(t)]^2\right)
dt, 
\end{multline}
which simplifies to
\begin{multline}
(***) = \frac{2|S|(|S|-1)}{n(n-1)}h_S(t) \, dt -
\frac{1}{2n(n-1)}\sum_{j\not=k\in S}h_{S\setminus\{j,k\}}
\, dt + \\
\frac{|S|}{n(n-1)}\sum_{k\in
S^c}\sum_{\alpha_k=0}^3\tr((\Psi^{\alpha_k}_S)^2) \, dt -
\frac{1}{2n(n-1)}\sum_{\substack{j\in S\\ k\in
S^c}}\sum_{\alpha_k=0}^3
\tr((\Psi^{\alpha_k}_{S\setminus\{j\}})^2)\, dt.
\end{multline}
After straightforward manipulations the expression further reduces to
\begin{multline}
(***) = \frac{2|S|(|S|-1)}{n(n-1)}h_S(t)dt -
\frac{1}{2n(n-1)}\sum_{j\not=k\in S}h_{S\setminus\{j,k\}}
dt + \\
\frac{2|S|}{n(n-1)}\sum_{k\in S^c}h_{S\cup\{k\}}dt -
\frac{1}{n(n-1)}\sum_{\substack{j\in S\\ k\in
S^c}}h_{S\setminus\{j\}\cup\{k\}}dt.
\end{multline}
Combining $(*)$, $(**)$ and $(***)$ then averaging over the realizations of the
Brownian motion yields the following system of coupled ODE's:
\begin{multline}
\frac{d\overline{h}_S(t)}{dt} = \frac{2|S^c|}{n(n-1)} \sum_{j\in
S} \overline{h}_{S\setminus \{j\}}(t) +
\left(\frac{2|S^c|(|S^c|-1)}{n(n-1)} + \frac{2|S|(|S|-1)}{n(n-1)} -2\right)\overline{h}_S(t) + \\
\frac{2|S|}{n(n-1)}\sum_{k\in S^c}\overline{h}_{S\cup\{k\}} -
\frac{1}{n(n-1)}\sum_{\substack{j\in S\\ k\in
S^c}}\overline{h}_{S\setminus\{j\}\cup\{k\}}.
\end{multline}
\section{Solutions of the purity ODE system}\label{ODEApp}
This appendix discusses solutions of the system of ODE's 
\begin{equation}\label{ode1}
 \frac{d\overline{h}_k}{dt}=\frac{k(n-k)}{n(n-1)}\:\left(2\overline{h}_{k-1}+2\overline{h}_{k+1}-5\overline{h}_k\right).
\end{equation}
We have investigated these equations numerically with initial conditions $\overline{h}_k = 1$, and found a logarithmic behavior in the ratio of scrambling times $\tau_* = t_*^{\kappa n} / t_*^1 \sim \log n$. Here, we will give a heuristic analytical argument for this behavior. For small values of $\frac{k}{n}$ and large $n$, the system in (\ref{ode1}) simplifies to
\begin{equation}\label{odea}
 \frac{d\overline{h}_k}{d\tau}=k\:\left(2\overline{h}_{k-1}+2\overline{h}_{k+1}-5\overline{h}_k\right),
\end{equation}
where $\tau=  t/n$. Define the tridiagonal matrix ${M}$ 
by ${M}_{k,k}=-5k$ and ${M}_{k,k\pm 1}=2k$ with $k=0,\hdots,n$
where the first row $k=0$ is all zeros.  Denote the eigenvalues of ${M}$ by $\lambda_j$. 
The eigenvector corresponding to $\lambda=0$ is simply $E_k^{(0)}=2^{-k}$. The eigenvalue
 problem ${M} E^{(\lambda)}=\lambda E^{(\lambda)}$ gives a set of recursive equations for $E_k^{(\lambda)}$
which have solutions of the form
\begin{eqnarray}
 E_k^{(\lambda)}=k 2^{-k}{}_2F_1\left(k+1,\frac{\lambda+3}{3},2,\frac{3}{4}\right),
\end{eqnarray}
where ${}_2F_1$ is the Gaussian hypergeometric function.
These eigenvectors blow up in the limit $k\to\infty$ unless $\lambda=-3j$ with $j$ a 
positive integer. The general solution to (\ref{odea}) in the limit 
$n\to \infty$ is given by
\begin{eqnarray}\label{sumj}
 \overline{h}_k(t)&=&\sum_{j=0}^\infty\:a_j\:E_k^{(j)}\:e^{-3j\tau}\nn\\
&=&2^{-k}a_0+\sum_{j=1}^\infty a_j\:k\:2^{-k}{}_2F_1\left(k+1,1-j,2,\frac{3}{4}\right) e^{-3j\tau}
\end{eqnarray}
At late time, the largest contribution comes form the zero eigenfunction, which selects $a_0 = 1$. We can get a sense for the relaxation time by examining the eigenfunction corresponding to the second eigenvalue, namely
the term with $j=1$. Direct evaluation of the hypergeometric function (which reduces to a polynomial in the above case) shows that the contribution of the $j=1$ eigenvalue is proportional to $2^{-k} k a_1 e^{-3\tau}$. Provided that the first correction qualitatively reflects the higher order corrections (which is does if $a_j$ decreases appropriately with $j$), we find $t_*^{k} \sim \log k$, so that $\tau_* \sim \log n$. 

\begin{figure}[t]
\centerline{a) \epsfig{figure=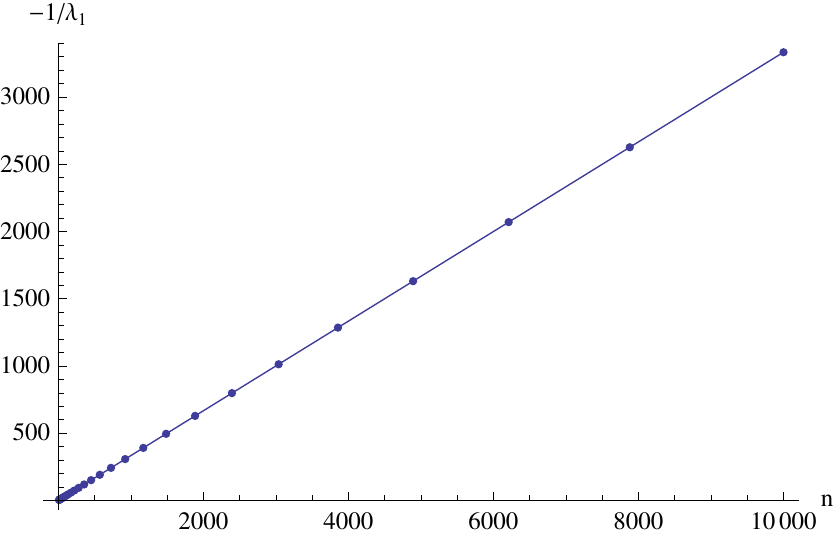, width=3in}  b) \epsfig{figure=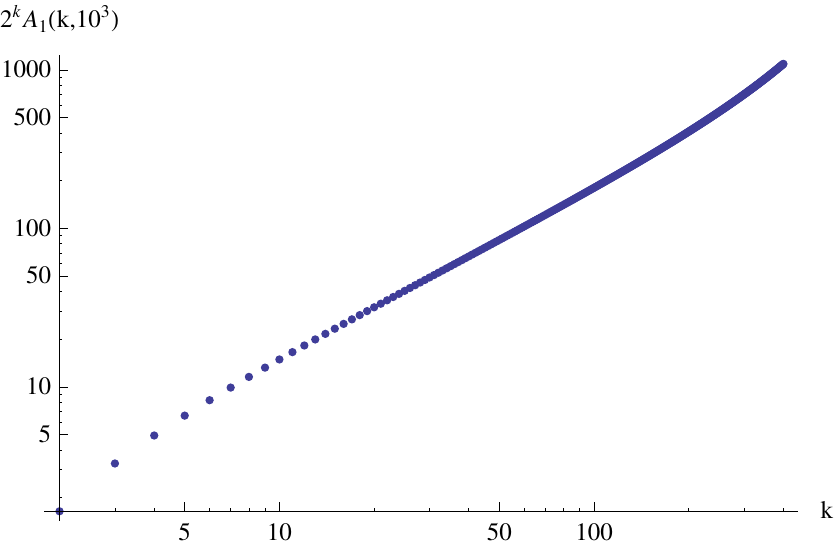, width=3in}}
\caption{
\label{eigen}
Largest nonzero eigenvalue $\lambda_1(n)$ and its corresponding eigenvector $A_1(k,n)$
 are computed numerically: a) the inverse of $\lambda_1$ with a negative sign is plotted 
as a function of $n$ suggesting 
$\lambda_1\simeq -\beta/n$ with $\beta\simeq2.99964$ b) the eigenvector $A_1(k,10^3)$ is
 multiplied by $2^k$ to shows the power law $k^\alpha$ with $\alpha \simeq1.346$.}
\end{figure}

Next, we turn to a numerical study of the eigenvectors for subsystems of larger $k/n$. 
Similarly, the solutions will have the general form
\begin{eqnarray}
\overline{h}_k(t)=\sum_{j=1}^n a_j e^{\lambda_j(n)t}A_j(k,n),
\end{eqnarray}
where the $\lambda_j(n)$'s are eigenvalues of the matrix ${B}$ (and therefore $k$-independent), 
and the $A_j(k,n)$ are the corresponding eigenvectors. It is only the largest nonzero 
eigenvalue and eigenvector that are
important for scrambling time. As can be seen in Figure \ref{eigen}, numerical results
 suggest that the largest nonzero eigenvalue $\lambda_1\simeq -3/n$ and its corresponding
 eigenvector $A_1(k)\sim 2^{-k}\:k^\alpha$ for $\alpha\sim O(1)$.

\section{$r$-body interactions and the BFSS matrix model}\label{sec:hypergraph}

Here, we revisit the Lieb-Robinson argument presented above for systems with $r$-body
 nonlocal interactions. The Hamiltonian for such systems has the form $H = \sum_X H_X$,
where the sum is over subsets of maximum size $r$ and $H_X$ acts on $\ox_{x \in X} \cH_x$. We will restrict our analysis to
 systems where $r$ is a constant, not a function of $n$. In analogy
 with the interaction graphs introduced in Section \ref{LR}, here the system
 can be represented by a hypergraph. Motivated by the fast scrambling conjecture, we focus on the BFSS matrix model as an example of a Hamiltonian with 
multi-body interactions, but the same type of argument can be used for other systems with $r$-body interactions, including those with complete $r$-uniform hypergraph Hamiltonians. The bosonic part of the Hamiltonian has the form
\bea\label{Ham_matrix}
H&=&\sum_a\tr\dot{M}^a\dot{M}^a+\sum_{a,b}\tr[M^a,M^b]^2\nn\\
&=&\sum_{a,i,j}\dot{M}^a_{ij}\dot{M}^a_{ji}+2\sum_{a,b,i,j,k,l}\left(M^a_{ij}M^b_{jk}M^a_{kl}M^b_{li}-M^a_{ij}M^a_{jk}M^b_{kl}M^b_{li}\right),
\eea
where the indices $a$ and $b$ range from 1 to 9 and the $M^a$ are $n$ by $n$ traceless Hermitian matrices. The degrees of freedom $M_{ij}^a$ are indexed 
by triples $(a,i,j)$ with $i\leq j$. The operators in the Hamiltonian have unbounded norm, so strictly speaking the Lieb-Robinson 
approach cannot be used. In this section we nonetheless proceed formally \emph{as if} the operators had bounded norm in order to determine whether the counting 
is consistent with a logarithmic signalling time.

The kinetic term $\dot{M}^a_{ij}\dot{M}^a_{ji}$ in (\ref{Ham_matrix}) is a single-body interaction, whereas the potential term is comprised of $4$-body interactions of the 
form 
$$
M_{ij}^a M_{jk}^b M_{kl}^a M_{li}^b
\qquad\mbox{and}\qquad
M_{ij}^a M_{jk}^a M_{kl}^b M_{li}^b.
$$
Repeating the same arguments as in the case of two-body 
interactions, for hypergraphs we find the inequality:\footnote{The discussion here parallels
 Appendix A of \cite{hastings:2006a}}
\begin{equation}\label{ineqr}
C_B(A,t) \leq  2\sum_{A\cap Z\neq \emptyset}\int_0^t ds\: C_B(Z,s) \|H_Z\|,
\end{equation}
where $Z$ is any multiset\footnote{A multiset is a generalization of a set in which members can be repeated.} of one or four degrees of freedom that has a nonzero contribution $H_Z$ to the Hamiltonian. $C_B(Z,t)$ itself is bounded from above by
\bea
C_B(Z,t) \leq  C_B(Z,0)+2\sum_{Z\cap Z'\neq \emptyset}\int_0^t ds\: C_B(Z',s) \|H_{Z'}\|.
\eea

At $t=0$ the operators $O_A$ in $A$ and $O_B$ in $y$ commute and therefore
\be
C_B(Z,0)\leq \left\{
  \begin{array}{l l}
    2\|O_B\|,  & \quad \text{for $Z\ni y$}\\
    0, & \quad \text{otherwise}.\\
  \end{array} \right.
\ee
Iterating the above inequality, one obtains
\begin{eqnarray}\label{r-body}
C_B(A,t) &\leq & 	 2\|O_B\|\left((2t) \sum_{\substack{Z:Z\cap A\neq \emptyset, Z\ni y}} \|H_Z\| +\frac{(2t)^2}{2!} \sum_{\substack{Z,Z':\\Z\cap A\neq\emptyset\\Z\cap Z'\neq\emptyset,Z'\ni y}}\|H_Z\| \|H_{Z'}\|+ \cdots\right).\nn\\
\end{eqnarray}
The contribution of each degree of freedom $(a,i,j)$ to the energy is bounded by
\begin{eqnarray}\label{singleen}
 \sum_{Z\ni (a,i,j)}\|H_Z\|&\leq&\big\|\dot{M}^a_{ij}\big\|^2+8\sum_{b,k,l}\left(\big\|M_{ij}^aM^b_{jk}M^a_{kl}M^b_{li}\big\|+\big\|M_{ij}^aM^a_{jk}M^b_{kl}M^b_{li}\big\|\right).
\end{eqnarray}
Note that the potential part of the above energy bound has $O(n^2)$ terms. We require the kinetic and the potential parts to be separately finite in the limit $n\to\infty$. One way to satisfy this is to introduce the following constraints:
\begin{eqnarray}\label{constrr}
 &&\|\dot{M}^a_{ij}\| \leq p\qquad \forall (a,i,j),\nn\\
&&\|H_X\|\leq \frac{c}{n^2}\qquad \forall X:|X|=4,
\end{eqnarray}
for positive constants $c$ and $p$. 

\begin{figure}
\begin{center}
\includegraphics[height=3.5cm]{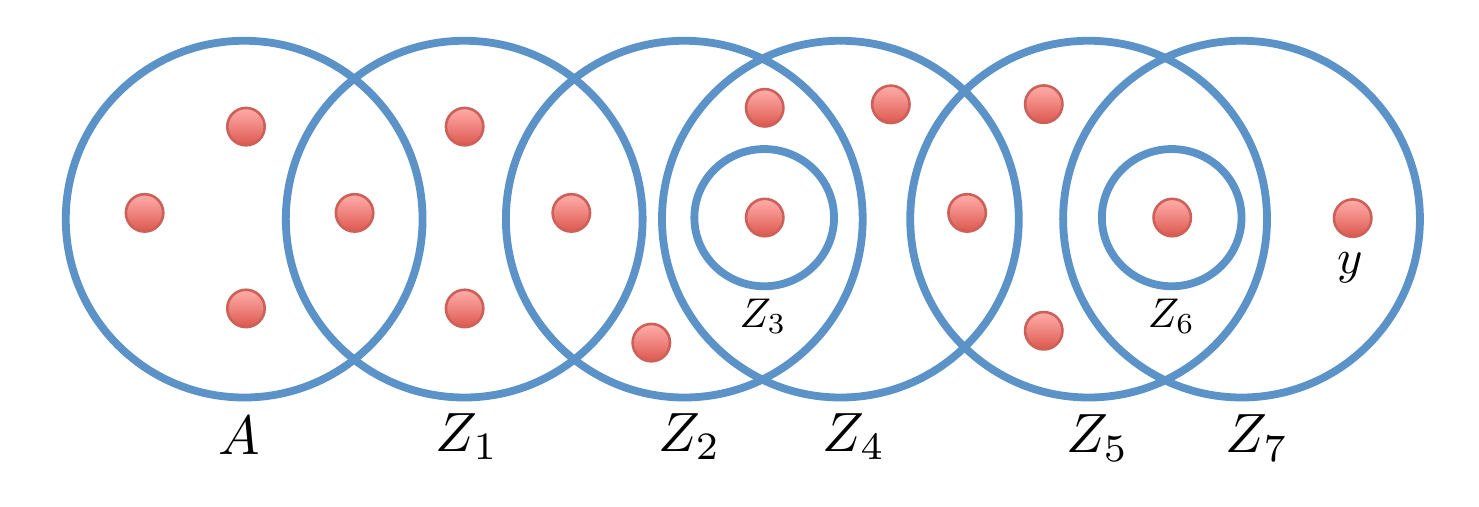}
\caption{The interaction hypergraph of the BFSS matrix model includes hyperedges that contain one or four vertices. The Lieb-Robinson bound in (\ref{r-body}) is found by summing over
a set of hyperedges that contain a path between $y$ and $A$. This figure illustrates a typical path connecting $y$ and $A$ with seven hyperedges.} \label{fig:4.body.counting}
\end{center}
\end{figure}

We are interested in finding an upper bound for the right hand side of (\ref{r-body}). This requires counting the number of terms in the $i^{th}$ sum in (\ref{r-body}).
Figure \ref{fig:4.body.counting} illustrates the type of subsets that correspond to the terms in the sum.
Using the constraints in (\ref{constrr}), the $i^{th}$ term can be bounded from above by
\bea\label{hypersum}
\sum_{\substack{Z_1,Z_2...,Z_i:\\Z_1\cap A\neq\emptyset\\\vdots\\Z_i\cap Z_{i-1}\neq\emptyset,Z_i\ni y}}\|H_1\|\cdots\|H_i\|\leq\sum_{k=0}^{i-1}\sum_{\substack{X_1,X_2...,X_{i-k}:\\X_1\cap A\neq\emptyset\\\vdots\\X_{i-k}\cap X_{i-k-1}\neq\emptyset,X_{i-k}\ni y}}\binom{i}{k} \:p^k\:\left(\frac{c}{n^2}\right)^{i-k},
\eea
where $k$ is the number of single-body multisets among $Z_1,\cdots,Z_i$ and 4-body multisets are denoted by $X$.  $\binom{i}{k}$ counts the number of ways of choosing $k$ of $i$ multisets to have only one degree of freedom.
Next, we focus on counting the number of terms in the sum on the right hand side of (\ref{hypersum}). Denote this number by $P_{i-k}$. 
If $p_{(j,j+1)}$ is the number of ways $X_j$ can intersect $X_{j+1}$, then
\bea
P_{i-k}\leq p_{(A,1)}\:p_{(1,2)}\:p_{(2,3)}\cdots p_{(i-k-1,i-k)}.
\eea
Notice that each four-body interaction term $M^{a_1}_{ij}M^{a_2}_{jk}M^{a_3}_{kl}M^{a_4}_{li}$ in the Hamiltonian has four indices $i,j,k$ and $l$ 
that run from $1$ to $n$. Fixing one degree of freedom fixes 
two of these indices, while fixing a second degree of freedom leaves only one index. Therefore, $p_{(j,j+1)}$ is order $n^2$ if $y\notin X_{j+1}$ 
and is order $n$ if $y\in X_{j+1}$.
Since $y$ has to belong to $X_j$ for some $j$, there are a maximum of $P=O\left(n^{2(i-k)-1}\right)$ nonzero terms in 
the sum (\ref{hypersum}). Plugging this result back in (\ref{hypersum}) gives
\bea
\sum_{\substack{Z_1,Z_2...,Z_i:\\Z_1\cap A\neq\emptyset\\\vdots\\Z_i\cap Z_{i-1}\neq\emptyset,Z_i\ni y}}\|H_1\|\cdots\|H_i\|&\leq& \frac{c'}{n}\sum_{k=0}^{i-1}\frac{i!\:c^i}{k!(i-k)!}\left(\frac{p}{c}\right)^k+O\left(n^{-2}\right)\nn\\
&=&\frac{c'}{n}\left((c+p)^i-p^i\right)+O\left(n^{-2}\right).
\eea
for some positive constant $c'$.
Now from (\ref{r-body}) we find the inequality
\bea
C_B(A,t)&\leq& \frac{2\|O_B\|\: c'}{n}\sum_{i=1}^\infty\frac{(2t)^i}{i!}\left((c+p)^i-p^i\right)+O\left(n^{-2}\right)\nn\\
&=&\frac{2\|O_B\|\: c'}{n}\left(e^{2(c+p)t}-e^{2pt}\right)+O\left(n^{-2}\right).
\eea
This finishes the ``proof'' that in the BFSS matrix model, signalling takes time at least $t_{signal} \geq O(\log n)$. Of course, we have really just proved the weaker statement that  a logarithmic lower bound holds for a related system with bounded operators in its Hamiltonian. It is therefore conceivable that this proof could be adapted to hold for the real BFSS Hamiltonian for all states in a low energy subspace. Alternatively, Lieb-Robinson bounds for lattice systems have been proved for some Hamiltonians containing unbounded operators~\cite{nachtergaele:2008a}. Similar techniques might be applicable to the matrix model.

\end{appendix}

\bibliographystyle{jhep}
\bibliography{scramble}

\end{document}